\documentclass[a4paper,fleqn,usenatbib]{mnras}

\usepackage{newtxtext,newtxmath}
\usepackage{lineno}

\usepackage[T1]{fontenc}
\usepackage{ae,aecompl}

\usepackage{graphicx}	
\usepackage{natbib, xcolor}
\newcommand{\novamon}{V959~Mon}
\newcommand{\swift}{{\sl Swift\/}}
\newcommand{\chandra}{{\sl Chandra\/}}
\newcommand{\suzaku}{{\sl Suzaku\/}}
\newcommand{\nustar}{{\sl NuSTAR\/}}
\newcommand{\fermi}{{\sl Fermi\/}}
\newcommand{\kms}{km\,s$^{-1}$}
\newcommand{\eps}{erg\,s$^{-1}$}
\newcommand{\epsc}{erg\,s$^{-1}$cm$^{-2}$}
\newcommand{\nh}{N$_{\rm H}$}
\newcommand{\nhism}{N$_{\rm H,ISM}$}
\newcommand{\nhint}{N$_{\rm H,int}$}
\newcommand{\nodata}{~$\ldots$~}%

\title[A non-radiative shock in nova \novamon]{X-ray evolution of the nova \novamon\ suggests a delayed ejection and a non-radiative shock}

\author[T. Nelson et al.]{Thomas Nelson,$^{1}$
Koji Mukai,$^{2,3}$\thanks{E-mail: Koji.Mukai@nasa.gov (KM)}
Laura Chomiuk,$^{4}$
Jennifer L.~Sokoloski,$^{5,6}$
\newauthor Justin D. Linford,$^{7}$
Thomas Finzell,$^{4}$
Amy J. Mioduszewski,$^{7}$
Michael P. Rupen,$^{8}$
\newauthor Jennifer Weston$^{9}$\thanks{Present address: Federated IT, 1201 Wilson Blvd, 27th Floor, Arlington, VA 22209, USA} and Raimundo Lopes de Oliveira$^{10,11}$
\\
$^{1}$Department of Physics and Astronomy, University of Pittsburgh, 3941 O'Hara St., Pittsburgh, PA 15260, USA\\
$^{2}$CRESST II and X-ray Astrophysics Laboratory, NASA/GSFC, Greenbelt, MD 20771, USA\\
$^{3}$Department of Physics, University of Maryland Baltimore County, 1000 Hilltop Circle, Baltimore MD 21250, USA\\
$^{4}$Center for Data Intensive and Time Domain Astronomy, Department of Physics and Astronomy, Michigan State University, East Lansing, MI 48824, USA\\
$^{5}$Columbia Astrophysics Laboratory, Columbia University, 550 West 120th St., New York, NY, USA\\
$^{6}$LSST Corporation, 933 N. Cherry Ave., Tucson, AZ 85721, USA\\
$^{7}$National Radio Astronomy Observatory, P.O. Box O, Socorro, NM 87801, USA\\
$^{8}$Herzberg Institute of Astrophysics, National Research Council of Canada, Penticton, BC, Canada\\
$^{9}$Green Bank Observatory, P.O. Box 2, Green Bank, WV 24944, USA\\
$^{10}$Departamento de F\'isica, Universidade Federal de Sergipe, Av. Marechal Rondon, S/N, 49000-000 S\~ao Crist\'ov\~ao, SE, Brazil\\
$^{11}$Observat\'orio Nacional, Rua Gal. Jos\'e Cristino 77, 20921-400, Rio de Janeiro, RJ, Brazil
}

\date{Accepted XXX. Received YYY; in original form ZZZ}

\pubyear{2020}

\begin{document}
\label{firstpage}
\pagerange{\pageref{firstpage}--\pageref{lastpage}}
\maketitle

\begin{abstract}
X-ray observations of shocked gas in novae can provide a useful probe of
the dynamics of the ejecta. Here we report on X-ray observations of the nova
\novamon, which was also detected in GeV gamma-rays with the \fermi\ satellite.
We find that the X-ray spectra are consistent with a two-temperature plasma
model with non-solar abundances. We interpret the X-rays as due to shock
interaction between the slow equatorial torus and the fast polar outflow
that were inferred from radio observations of \novamon. We further propose that
the hotter component, responsible for most of the flux, is from the reverse
shock driven into the fast outflow. We find a systematic drop in the column
density of the absorber between Days 60 and 140, consistent with the
expectations for such a picture. We present intriguing evidence for a delay of
around 40 days in the expulsion of the ejecta from the central binary.
Moreover, we infer a relatively small (a few times 10$^{-6}$ M$_{\odot}$)
ejecta mass ahead of the shock, considerably lower than the mass of 10$^4$ K
gas inferred from radio observations. Finally, we infer that the dominant X-ray
shock was likely not radiative at the time of our observations, and that the
shock power was considerably higher than the observed X-ray luminosity. It is
unclear why high X-ray luminosity, closer to the inferred shock power, is never
seen in novae at early times, when the shock is expected to have high enough
density to be radiative.
\end{abstract}




\section{Introduction}
\label{intro}

Nova eruptions are the most common class of stellar explosion in
the universe. They occur when a white dwarf gains enough material
from a mass-losing binary companion to trigger a thermonuclear
runaway in the accreted shell \citep{CNII}.  This releases a large
amount of energy (10$^{44}$--10$^{46}$ erg) through nuclear burning
and subsequent decays of radioactive nuclei, and drives the expulsion
of much, if not all, of the shell into the circumbinary environment.
Although novae are most commonly discovered in the optical as a result
of their dramatic increase in visual brightness, they are truly
panchromatic events, showing complex, inter-related evolution at
all wavelengths from radio to gamma-rays.  Each regime generally
provides just one view of the eruption; to truly capture the physics
of the explosion and ejection process in detail, a synthesis of
observations at many wavelengths is required. 

X-ray emission is frequently observed in novae at some point during
the eruption, and has two distinct origins.  The first type of emission
is typically observed in the hard (1--10 keV) energy band, and is thought
to originate in high-temperature, optically-thin, shocked gas. These shocks
form through interaction with the dense wind of a red giant companion in
the case of nova eruptions occurring in symbiotic systems, such as RS Oph
\citep{Sokoloski06}, V407 Cyg \citep{Nelson12}, and V745 Sco
\citep{Orio15}. However, the majority of nova eruptions occur in
cataclysmic variables, in which the mass donors are late type stars
on or near the main sequence and do not have significant winds.
In such cases, the X-rays originate in internal shocks in the nova
ejecta as faster outflows sweep up and shock some earlier, slower
stage of mass loss \citep{O'Brien94, Metzger14}.

The hard X-rays detected in novae are often highly absorbed at early times.
In some well-studied cases, the absorbing column towards the X-ray emitting
region has been observed to decline over time, presumably from the expansion
of the outer parts of the ejecta.  The evolution of \nh\ in these cases
can be used to constrain the mass of the nova ejecta external to the
shocked region \citep{Balman98, Mukai01}.  The temperature of the post-shock
gas reveals information about the velocity differential between the two
flows via the strong shock conditions (see Section 4, below). Hard
X-rays are common in novae \citep[see e.g. ][]{Schwarz11}, and have
been proposed as the origin of some hard X-ray transients observed
toward the Galactic center \citep{Mukai08}.

The second type of X-ray emission observed in novae originate
in the photosphere of the nuclear shell-burning white dwarf.
This is the optically thick, blackbody-like ``supersoft'' emission
that is characterized by effective temperatures in the range 20--100 eV.
This component becomes observable only after the nova ejecta have expanded
sufficiently to become optically thin to soft X-rays.
The supersoft component has luminosities of order 10$^{36}$
to 10$^{38}$ \eps, several orders of magnitude higher than that of
the harder shock emission, and provides a direct probe of the white dwarf.
The flexible scheduling of the Neil Gehrels {\it Swift}
Observatory (hereafter \swift) has enabled detailed studies of the
supersoft phase of a large number of novae in recent years
(see, e.g., \citealt{Ness07,Schwarz11} and references therein).
However, in this paper, we concentrate on the harder shock emission
in an attempt to improve our understanding of the mass ejection
processes in novae.

The properties of the nova ejecta that are elucidated by X-ray
observations can be considered in tandem with data from other
wavelengths to build up a complete picture of the mass ejection
process during eruption.  While X-ray emission reveals the temperature
of the post shock region, and by extension the velocity difference
of the interacting media, optical spectroscopy provides constraints
on the velocity of the fastest ejecta \citep[see e.g.][]{Diaz10}.
Radio observations directly probe the ionized gas in the nova shell,
and can trace the density structure and expansion history of the
ejecta \citep{Seaquist77}. They can also reveal the presence of
accelerated particles via non-thermal emission \citep{Weston16}.
Finally, \fermi\ observations have revealed that nova eruptions
can lead to rapid, efficient particle acceleration and the emission
of GeV gamma-rays during the first few weeks of the onset of eruption
\citep{Ackermann14, Cheung16}, providing further information on
the nature of shocks and mass loss in novae.  

\subsection{\novamon}

\novamon\ is one of the novae that have been detected as GeV gamma-ray
transients with the Large Area Telescope (LAT) instrument onboard
the \fermi\ satellite \citep{Ackermann14}.  Interestingly, the gamma-ray
transient was not immediately identified with a nova, as its position
was too close to the Sun for follow-up at most other wavelengths.
We take the time of the \fermi\ transient discovery, 2012 June 19.0, 
or MJD 56097, as $t_{0}$ for this eruption\footnote{Note that time of the
first gamma-ray discovery in \citet{Ackermann14} is three days earlier
than the date initially reported by \citet{Cheung12a}, or 2012 June 22.
This means that our $t_{0}$ definition differs from some studies published
prior to 2014.}. The association with a nova was not made until
2012 August 9 (Day 51) when \novamon\ was discovered in the optical
\citep{Fujikawa12, Cheung12b}.  An intensive multi-wavelength campaign
was initiated in response to the discovery of the nova in the optical,
and included radio, infrared, optical, UV and X-ray observations.

Both \citet{Ribeiro13} and \citet{Shore13} discussed high-resolution
optical spectroscopy of \novamon.  Based on the similarity of the
optical spectra to those of the nova V382~Vel, Shore et al.
classified \novamon\ as an oxygen-neon (ONe) nova that was
first observed in the optical well after maximum light.  Ribeiro et al.
were able to model the highly-structured emission lines by assuming
a bipolar morphology for the ejecta viewed at high inclination
($\sim$82$^{\circ}$ $\pm$ 6$^{\circ}$). The maximum expansion velocity
of the ejecta is 2400$^{+300}_{-200}$ \kms. Both works show that the
optical emission line velocities were relatively stable between
days 55 and 190, with no indications of drastic velocity changes
or of emergence of a new component.

\citet{Chomiuk14a}, \citet{Linford15} and \citet{Healy17} presented a
series of high-resolution radio images of \novamon\ taken using Karl
G. Jansky Very Large Array (VLA), Very Long Baseline Array (VLBA),
and enhanced Multi Element Remotely Linked Interferometer Network
(e-MERLIN), and use these to trace the evolution of the nova ejecta
over the course of the eruption.  The ejecta were
spatially resolved in the radio starting on Day 91, and were observed
to evolve in both size and shape over the course of the eruption.
The images reveal the presence of an asymmetry that rotated by 90
degrees over the course of the eruption.  \citet{Chomiuk14a} interpret
this structure as follows: at early times, a common envelope was formed
around the binary by mass loss preferentially in the plane of the binary.
Some time later, a faster outflow began, driving mass loss primarily
in the polar direction.  Since this material was moving faster,
it quickly became more spatially extended than the common envelope
structure, and dominated the radio morphology in the image obtained
on Day 126.  At much later times the fast wind dropped in density,
leaving the denser common envelope as the primary source of surface
brightness of the nova remnant in an image obtained on Day 615.
In this scenario, the secondary star plays a key role in ejecting
the shell from the central binary.  \citet{Healy17} observe a similar
evolution in morphology in images obtained with the e-MERLIN array.
\citet{Linford15} used the VLA dataset in conjunction with optical
spectroscopy to derive a distance to the nova by modeling the expansion
of the ejecta.  They find a best distance to the nova of 1.4 $\pm$ 0.4 kpc;
we adopt 1.4 kpc as the distance throughout this paper in deriving
the emission measure and the luminosity.

\citet{Page13} presented the overall evolution of the \novamon\ eruption
in X-rays and UV as observed with \swift.  Two distinct X-ray emitting
components were identified based on the very different evolution of flux
above and below 0.8 keV.  The harder component, presumed to be emission
from shocked gas, dominated until Day 162.  At that time, a softer
component emerged in the spectrum that was identified as supersoft
emission from the white dwarf photosphere.  A period of 7.1 hrs was
detected in the periodogram of the X-ray, UV and optical light curves
that the authors identify as the orbital period of the system.  The
presence of phased modulation from X-rays to near-IR emission is
interpreted as the presence of a disk rim bulge viewed at moderately
high inclination, consistent with the spectral modeling results presented
by \citet{Ribeiro13}.  In addition, Figure 1 of \citet{Page13} shows that
\novamon\ declined smoothly in the optical and UV throughout the period
covered by their observations, Day 51 through 259.

\citet{Peretz16} presented an analysis of
two high-resolution grating spectra of \novamon\ obtained with
the \chandra\ observatory on Day 85 and and on Day 167 of the eruption.
The authors observed emission lines consistent with the presence
of shocked plasma in both observations, and evidence of continuum
emission from the white dwarf surface in the later spectrum.
They also infer highly non-solar abundances in the X-ray emitting
material, most notably of neon, magnesium and aluminum.  The authors
claim that the X-rays originate in high density clumps in the ejecta,
based on density diagnostics that use emission lines of He-like ions.

\begin{figure}
\centering
\includegraphics[width=3in]{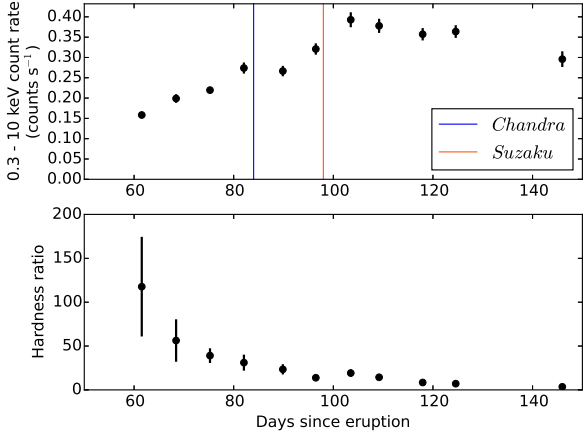}
\caption{\swift\ XRT 0.3--10 keV count rate ({\it upper panel}), and
hardness ratio ({\it lower panel}), defined here as the ratio hard/soft
of the count rates in the 1--10 keV (hard) and 0.3--1 keV (soft) bands.
The blue and red lines indicate the times of the \chandra\ and
\suzaku\ observations, respectively.}
\label{fig:swiftcurve}
\end{figure} 

In this paper, we focus on the optically thin X-ray emission observed
with \swift, \chandra, and \suzaku\ from 61 to 155 days after the initial
gamma-ray discovery, with the goal of probing the dynamics and mass
of the ejecta.  We reanalyze the \chandra\ spectrum from Day 85
presented in \citet{Peretz16} in order to inform the analysis of the
\suzaku\ and \swift\ observations.  The data presented here
were all obtained prior to the emergence of the white dwarf photosphere
on Day 152. For a detailed study of the supersoft X-ray emission, see
both \citet{Page13} and \citet{Peretz16}.

\section{Observations and Data Reduction}

X-ray observations of \novamon\ were obtained with the \swift, \suzaku,
and \chandra\ satellites.  Here we provide details of the observations
and the data reduction procedures for each satellite.

\label{data}
\subsection{\swift}
The \swift\ satellite began monitoring \novamon\ on 2012 August 19, shortly
after the announcement of the optical discovery of the nova.   Observations
using the X-ray Telescope (XRT) were initially carried out at a roughly
weekly cadence until the discovery of supersoft emission from the nova
in the data obtained on 2012 Nov 28 (Day 162), at which point a daily
observing campaign was initiated.  Since the focus of this paper is
the hard X-ray emission, we concentrate on the observations taken
through 2012 Nov 11 (Day 146; see Table~\ref{tab:obslog} for details
of the observations).  

\begin{figure*}
\centering
\includegraphics[width=5in]{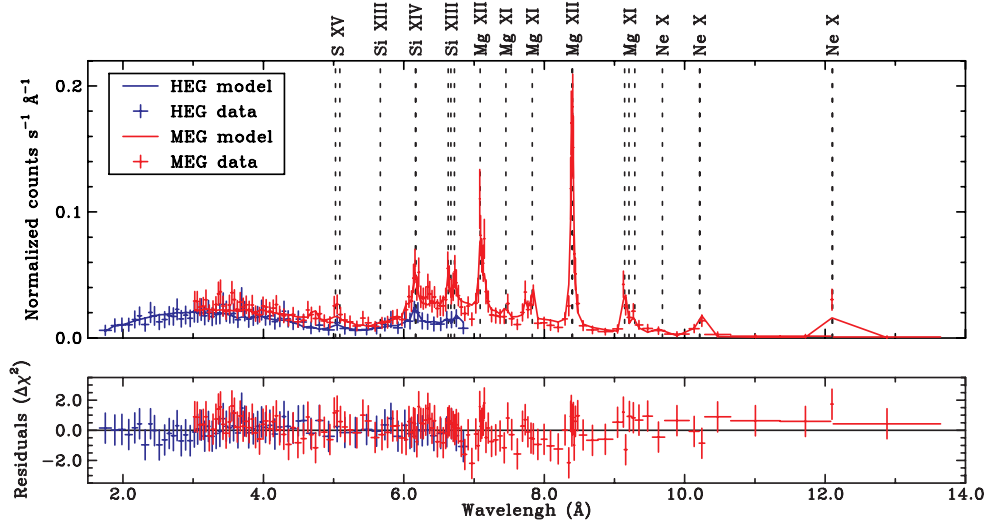}
\caption{\chandra\ HETG data from Day 85, with best-fit model two temperature
{\tt bvapec} model shown as solid lines. The data are the summed 1st order HEG
(blue) and MEG (red) spectra, grouped to have a minimum of 25 counts per
bin.  Residuals of the model fit (see Table 2 for parameters) are shown in
the lower panel. The positions of strong emission lines, corrected for the
best-fit blueshift of 770 \kms, are shown as dashed lines.}
\label{fig:chandra_spec}
\end{figure*}

All of the XRT data included in this paper were obtained in photon
counting (PC) mode, with exposure times ranging from 1880 to 5880 seconds.
We created spectra for each observation using xselect v2.4b.  Source photons
were extracted from a circular region of radius 20 pixels ($\sim$47 arcsec)
centered on the nova, while background events were extracted from a larger
circular region located off the source.  The source spectra were binned
the to have a minimum of one count per bin. We used the {\tt xrtmkarf}
tool to create ancillary response files (ARFs) for each spectrum,
correcting for dead columns and pixels using the exposure map included
with the data from the archive.   The source count rate in all observations
was below 0.4 cts\,s$^{-1}$, so no additional corrections were made for pile-up.
Finally, we downloaded the appropriate response matrix file (RMF) from
the calibration database, in this case swxpc0to12s6\_20110101v014.rmf.
In our spectral fitting of \swift\ data, we binned the data to have a
minimum of one count per bin, and used the C-statistic to determine best
fit model parameters in the energy range 0.3--10 keV.

\begin{table}
\centering
\caption{Observation details}
\begin{tabular}{cccc}
\hline
Date & Obs ID & $t_{\rm exp}$ & Time since $t_{0}$ \\
UT & & (s) & (days) \\
\hline
\multicolumn{4}{c}{\underline{\swift}}\\
2012 Aug 19 & 00032529001 & 5799 & 61.2  \\ 
2012 Aug 26 & 00032529002 & 2033 & 68.4  \\
2012 Sep 02 & 00032529003 & 5876 & 75.1  \\ 
2012 Sep 09 & 00032529004 & 1704 & 82.0  \\ 
2012 Sep 16 & 00032529005 & 2017 & 89.8  \\ 
2012 Sep 23 & 00032529006 & 2001 & 96.4  \\ 
2012 Sep 30 & 00032529007 & 1879 & 103.5  \\ 
2012 Oct 06 & 00032529008 & 2023 & 109.2  \\ 
2012 Oct 14 & 00032529009 & 1923 & 117.9  \\ 
2012 Oct 21 & 00032529010 & 1924 & 124.5  \\ 
2012 Nov 11 & 00032529011 & 1924 & 146.0  \\ 
\hline
\multicolumn{4}{c}{\underline{\chandra}}\\
2012 Sep 12 & 15495 & 24459 & 85  \\
\hline
\multicolumn{4}{c}{\underline{\suzaku}}\\
2012 Sep 25 & 907002010 & 46886 & 98  \\
\hline
\end{tabular}
\label{tab:obslog}
\end{table} 

\subsection{\chandra}

In response to the discovery of bright X-ray emission from a nova detected
as a  gamma-ray source, a directors discretionary time (DDT) observation
of \novamon\ was carried out with the \chandra\ satellite on 2012 Sept 12
(MJD 56182, Day 85 of the eruption; see
Table~\ref{tab:obslog}).  The total exposure time was 24.5 ks,
and the observation was carried out using the high energy transmission
grating (HETG) and the ACIS-S camera. The HETG instrument is comprised
of two gratings, the High Energy Grating (HEG) and Medium Energy Grating
(MEG), that in combination provide high spectral resolution over the
wavelength range 1.5--30 \AA\ (corresponding to photon energies of
$\sim$0.4-8.2 keV).  Preliminary results were reported by
\citet{Ness12}, who noted the presence of blue-shifted emission lines
and probably non-solar chemical abundances.  A more detailed analysis
of the \chandra\ spectrum was published previously by \citet{Peretz16},
as noted earlier.

We chose to reanalyze this \chandra\ spectrum in tandem with our
exploration of the \suzaku\ and \swift\ datasets.  We re-processed
the data downloaded from the archive using the {\tt chandra\_repro}
script and CIAO version 4.7.  The processing script created new level
2 event files, and from that extracts the level two PHA files that
contain the spectra.  It also creates the response matrix (RMF) and
ancillary response (ARF) files required for spectral modeling.
Each spectrum was binned by a factor of two in channel space to increase
signal to noise but maintain the energy resolution of the instrument.
We fitted the four spectra ($\pm$1st orders for both HEG
and MEG) independently, and used the C-statistic \citep{Cash79}
to obtain best-fit model parameters and associated uncertainties.
However, for plotting purposes, we co-added the +1 and $-$1 orders using
the CIAO script ({\tt combine\_grating\_spectra}) to increase the
signal to noise in each spectral bin.

\subsection{\suzaku}

We requested a DDT observation of \novamon\ with the
\suzaku\ satellite, which was approved and carried out  on 2012
September 25 (MJD 56195, or Day 98 of the eruption; see
Table~\ref{tab:obslog}).  The total exposure time was 46.9 ks.
No source was detected with the HXD instrument, so we focus on
the data obtained with the X-ray Imaging Spectrometer (XIS) in
the 0.3--10 keV energy range. All three functioning XIS units were
operated in the full-window imaging mode. We extracted the source
spectra from a circular region with a 3.5\arcmin~radius centered
on \novamon, and background spectra from annular regions also centered
on the source, with inner radius of 4\arcmin~ and outer radius 6.5\arcmin.
We created response files using the {\tt xisrmfgen} and {\tt xisarfgen}
ftools, using the version 20120719 contamination files. Given
the larger number of counts collected, we used the $\chi^{2}$ statistic
to obtain best-fit model parameters and uncertainties.

\section{X-ray evolution of \novamon\ as observed with \swift, \chandra, and \suzaku}
\label{xanalysis}

The 0.3--10 keV count rate and hardness ratio for the \swift\ observation
sequence are plotted in Figure~\ref{fig:swiftcurve}, which reveals an X-ray
evolution typical of novae weeks to months after the start of the eruption.
The hardness ratio is defined here as the hard/soft count rate ratio above
and below 1 keV. The X-ray emission becomes both brighter and softer
with time through Day 100 or so, then levels off
(Figure~\ref{fig:swiftcurve}).

Although monitoring of \novamon\ with
\swift\ provides a useful global view of the evolution of the X-ray
emission, the short exposures and small number of collected photons
means that some of the details of the X-ray emitting region,
particularly those revealed by emission lines, are missed.
The abundances of nova ejecta are known to be highly non-solar, with
enhancements in CNO cycle elements from nuclear burning and in Ne if
the underlying white dwarf is of the ONe subtype
\citep[see e.g. ][and references therein]{Helton12}.  Furthermore,
the evolving nature of the eruption can lead to non-equilibrium
ionization effects where emission line ratios have different values
than those expected for plasmas in collisional ionization.  To explore
these details, we make use of the two deeper observations of \novamon\ that
were obtained with the \chandra\ and \suzaku\ satellites.

The cross-calibration uncertainties among these observatories
are small and well-understood, thanks to the efforts of the
International Astrophysical Consortium for High Energy
Calibration (IACHEC)\footnote{https://iachec.org/}. Cross-calibration
issues can generally be ignored in combined analyses of data from these
unless the observations are all well-exposed and have small statistical
errors so that differences become evident, which is not the case here.
We therefore infer that any disagreements in the fit results are due
to statistical limitations of the data, source variability, or our
limited understanding of the underlying physics, as reflected in our
choice of spectral models.

In this section, we present our spectral analysis of the data obtained
with all three satellites.  We first revisit the \chandra\ spectrum
presented in Peretz et al.  We then use the insights from the
\chandra\ modeling to analyze the \suzaku\ spectrum obtained on Day 98.
Finally, we analyze the set of \swift\ spectra using the abundances
found from the analysis of the two deep spectra.  Given the presence
of emission lines in the \chandra\ and \suzaku\ spectra, we modeled
all spectra in xspec v12.8.0m using the APEC suite of thermal,
collisional plasma models.  We used the {\tt tbabs} model for
foreground absorption assuming the photoionization cross-section
values of \citet{Verner96}. In addition, we initially assume the
the abundances of \citet{Wilms00} for the absorber. This assumption
is appropriate for the interstellar medium, but as we show below,
we detect significant absorption from parts of the nova ejecta.
We discuss the consequence of an alternative assumption regarding
the absorber composition, perhaps more appropriate for this situation,
in Section\,\ref{ejectamass}.

\subsection{Revisiting the Day 85 \chandra\ spectrum}

The \chandra\ spectrum obtained on Day 85 is dominated by emission lines
of hydrogen- and helium-like Si and Mg, and hydrogen-like Ne.
\citet{Peretz16} presented an analysis of this \chandra\ spectrum and
found an acceptable fit to the data with a two-temperature collisional
ionization equilibrium (CIE) plasma model
with highly enhanced abundances of metals, including Ne, Mg and Al.
The emission lines were blueshifted by $850_{-145}^{+75}$ \kms\ in the
low temperature component, but not in the hotter plasma (which were
frozen to zero).  Finally, both plasmas showed broadened lines, with
a FWHM velocity of 676$_{-70}^{+80}$ \kms; this value was assumed to
be the same in both temperature components.

We modeled the \chandra\ data utilizing the entire wavelength range of
the HEG (1.5--15 \AA; 0.83--8.3 keV), and a subsection of the MEG range
(2--20 \AA; 0.62--6.2 keV) as there is very little signal at longer
wavelengths.  In our fit, we used the same model ({\tt tbabs*(bvapec+bvapec)})
as \citet{Peretz16}, with a few key differences.  First, we assume abundances
for He, C, N and O determined from optical spectroscopy of
\novamon\ by \citet{Tarasova14}, and keep these values fixed.
Second, we allow the line broadening
of the two components to vary freely.  Finally, we assume a single velocity
shift for the two components, but compare our results with those of Peretz
et al. below.

In Figure~\ref{fig:chandra_spec}, we show the combined 1st order HEG and MEG
spectra with our best-fit model and residuals.  The resulting model
parameters, shown in Table~\ref{tab:spec_results}, are broadly compatible
with the findings of \citet{Peretz16}.  The two plasma temperatures are 3.7
and 0.64 keV, which are slightly lower values than those presented by
Peretz et al. (4.5 and 0.8 keV, respectively).
The absorbing column towards the X-ray emitting region is
(3.2 $\pm$ 0.2) $\times$ 10$^{22}$ cm$^{-2}$. The abundances of Ne, Mg, Al, Si
and S are all strongly enhanced relative to reference values (see
Table~\ref{tab:spec_results} for values).  The abundance of Fe is
poorly constrained by the spectra, which is not surprising given the
lack of strong Fe lines in the data,
and we find only an upper limit on Fe/Fe$_{\odot}$ of $<$1.5 at the 90\%
confidence level.  We find line broadening of 1370 and 510 \kms\ for
the 3.7 and 0.6 keV plasmas, respectively.  Finally, our best-fit model
indicates a blueshift in the line positions of 771$^{+71}_{-66}$ \kms,
which is within the uncertainty range found by \citet{Peretz16} for the
lower temperature component. We note that the fit statistic found when the
blueshift of the 3.7 keV component is fixed to zero (as in Peretz et al.)
is not significantly different to our best-fit, single velocity shift model,
and none of the other model parameters change within the uncertainties.
We take this to mean that there is a degeneracy in the line centroid and
the line width, when fitting two-component spectral models to the
\chandra\ grating data of this nova.  The data tell us that the two
components have different kinematic signatures, but we cannot tell
whether the difference is in the velocity shift or in line width.

\begin{table}
\centering
\caption{Best-Fit Model Parameters for \chandra\ and \suzaku\ Spectra.
Model is of form {\tt tbabs*(bvapec+bvapec)} with redshift and abundance
set equal for both bvapec components.}
\begin{tabular}{lcc}
\hline
&  \chandra & \suzaku \\
\hline
\nh\ (10$^{22}$ cm$^{-2}$) & 3.2 $\pm$ 0.2 & 1.44$^{+0.06}_{-0.05}$ \\
\\
blueshift (\kms) & 771$^{+71}_{-66}$ & 771$^a$\\
\\
kT$_{1}$ (keV) & 3.7$^{+0.5}_{-0.4}$  & 3.9 $\pm$ 0.1\\
norm$_{1}^{b}$ & 0.005 $\pm$ 0.002 & 0.009 $\pm$ 0.0003\\
velocity width (\kms) & 1362$^{+315}_{-259}$ & 1362$^a$ \\
\\
kT$_{2}$ (keV) &  0.64$^{+0.10}_{-0.05}$ & 0.311$_{-0.008}^{+0.009}$\\
norm$_{2}^{b}$ & 0.0010$_{-0.0006}^{+0.0007}$ & 0.0024 $\pm$ 0.0003\\
velocity width (\kms) & 515$^{+87}_{-77}$ & 515$^a$ \\
\\
He/He$_{\odot}$ & 1.5 & 1.5 \\
C/C$_{\odot}$ & 1 & 1 \\
N/N$_{\odot}$ & 33 & 33 \\
O/O$_{\odot}$ & 9.2 & 9.2 \\
Ne/Ne$_{\odot}$ & 207$_{-93}^{+323}$ & 19$^{+3}_{-2}$ \\
Mg/Mg$_{\odot}$ & 55$_{-21}^{+73}$ & 20 $\pm$ 2\\
Al/Al$_{\odot}$ & 51$^{-23}_{+70}$ & 25$^{+7}_{-6}$\\
Si/Si$_{\odot}$ & 6$^{+8}_{-3}$ & 2.3 $\pm$ 0.5\\
S/S$_{\odot}$ & 4$^{+5}_{-2}$ & 1.9$\pm$ 0.4\\
Fe/Fe$_{\odot}$ & $<$1.5 & 0.17$\pm$ 0.05\\
\\
XIS 1 normalization  & \nodata & 1.0\\
XIS 0 normalization &  \nodata & 1.00 $\pm$ 0.01\\
XIS 3 normalization &  \nodata & 1.01 $\pm$ 0.01 \\
\\
$F_{0.3-10}$ (10$^{-11}$ \epsc)$^{c}$ & 1.70$^{+0.04}_{-0.67}$ & 1.61 $\pm$ 0.02 \\
$L_{0.3-10}$ (10$^{34}$ \eps)  & 1.4 $\pm$ 0.2 & 1.07 $\pm$ 0.02 \\
& & \\
\hline
C-stat (\chandra) or $\chi^{2}$ (\suzaku) & 5799 & 3919\\
D.O.F & 9182 & 2684\\
\hline
\multicolumn{3}{l}{$^a$Velocity shift fixed to \chandra\ value.}\\
\multicolumn{3}{l}{$^b$norm = $\frac{10^{-14}}{4\pi D^{2}} \int n_{e} n_{i} dV$}\\
\multicolumn{3}{l}{$^c$We used the {\tt energies} command in xspec to extend the}\\
\multicolumn{3}{l}{energy coverage of the HETG response down to 0.3 keV.}\\
\multicolumn{3}{l}{All quoted uncertainties are 90\% confidence intervals.}\\
\end{tabular}
\label{tab:spec_results}
\end{table}

The lower temperature plasma component in the model appears to be necessary
to explain the ratios of H- to He-like lines in the spectrum (particularly
for Mg) which cannot be reproduced by a single temperature model.
Non-equilibrium ionization models can in principle
also account for H-like to He-like line ratios that depart from
the values expected in CIE models (see the case of V407~Cyg;
\citealt{Nelson12}). However, the density we infer for the X-ray
emission region makes this alternative interpretation unlikely,
as we discuss in subsection\,\ref{density}.

\begin{figure*}
\centering
\includegraphics[width=5in]{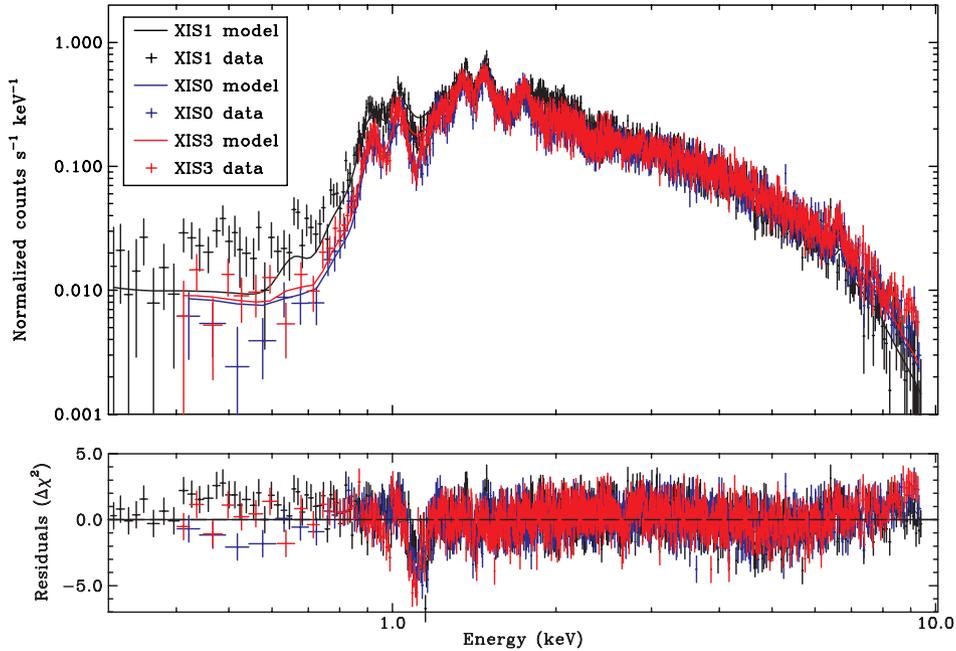}
\caption{\suzaku\ XIS data from Day 98, with best-fit model shown
as solid lines. Residuals of the model fit as shown in the lower panel.
The parameters of this model are shown in Table~\ref{tab:spec_results}.}
\label{fig:suzaku_spec}
\end{figure*}

\subsection{\suzaku\ view on Day 98}
\label{suzakuspec}

To model the \suzaku\ spectra, we used the 0.3--10 keV range for
the XIS1 instrument, and 0.4--10 keV for XIS0 and XIS3. Our starting point
was the best-fit model found for the \chandra\ data.  Since
the line-broadening derived from the model fits to the HETG spectra
is less than the instrumental spectral resolution of the \suzaku\ XIS CCDs,
we fixed the line broadening to those found for the \chandra\ spectra
when fitting the XIS data. We also fixed the blueshift value of the models
to $v$ = 771 \kms, the best-fit value found in modeling the \chandra\ data.
The resulting best-fit parameters are shown in Table 2, and the data,
model and residuals are shown in Figure~\ref{fig:suzaku_spec}.

The temperature of the hotter component in the two-temperature CIE model
(column 3 in Table 2) is 3.9 $\pm$ 0.1 keV, in agreement with the value
found for the \chandra\ spectrum within the uncertainties. In contrast,
a lower temperature of 0.32 $\pm$ 0.01 keV is found for the second component.
The normalizations, and hence emission measures, of both temperature
components are slightly higher than those found in the best-fit
\chandra\ model.  Given that the overall flux is slightly lower,
this is most likely an effect of the lower best-fit abundances in this model,
which are
lower than in the \chandra\ model, and have much smaller uncertainties.
This difference is most notable for neon, where we find
Ne/Ne$_{\odot}$ = 19$^{+3}_{-2}$, and for iron, where we find
Fe/Fe$_{\odot}$ = 0.17 $\pm$ 0.05.  Finally, the absorbing column
attenuating the X-ray emission is also lower than in the \chandra\ fit,
with \nh\ = 1.4 $\times$ 10$^{22}$ cm$^{-2}$.  This results in a lower
unabsorbed flux and inferred luminosity. 

The fit of this model is by no means perfect; the reduced $\chi^{2}$ value
is 1.46, and clear residuals are seen in the fit to the data.  The best-fit
model underestimates the \suzaku\ data at energies  $>$8 keV.  We note
that the \novamon\ region of the X-ray sky is somewhat crowded
\citep{Evans14} and it is possible that the excess is due to a
highly absorbed source within the 210" radius extraction region.
There is also
a significant negative residual ($\sim$6--7$\sigma$) at 1--1.1 keV. There
are several lines of Fe
and a line of helium-like Ne that is strong
in plasmas with kT of around 0.3 keV.
Given that
the best-fit Fe abundance is very low, the discrepancy is likely
to be due to the Ne line.
Finally, the apparent detection of signal below 0.7 keV
is entirely consistent with the low-energy tail in the CCD response to
higher energy photons, due to incomplete collection of charges. This is
a feature commonly seen in heavily absorbed sources, and
is known to be less well calibrated than the main part of the detector
response. We nevertheless show the data down to 0.3 (for XIS1)/0.4 (for
XIS0 and XIS3) keV, to demonstrate the absence of photons from
\novamon\ at these energies, such as the supersoft component or
strong low-energy emission lines.

\subsection{X-ray evolution observed with \swift\ is driven by evolving absorption}

We further investigated the X-ray evolution of \novamon, already seen
in the plot of 0.3--10 keV count rate and hardness ratio
(Figure~\ref{fig:swiftcurve}), by fitting models informed by our
analysis of \chandra\ and \suzaku\ spectra to the XRT spectra.
We first attempted to fit a two-temperature plasma model of the same
form as those in Table~\ref{tab:spec_results} to each XRT spectrum,
fixing the elemental abundances to those found for \chandra\ or \suzaku.
However, these model fits were unstable and parameter uncertainties
could not be derived.  Given that the higher temperature component
dominates the flux of the best-fit models for both the \chandra\ and
\suzaku\ spectra, we decided to model the X-ray emission in the XRT
spectra using single temperature model ({\tt tbabs*bvapec}), assuming
the best-fit \suzaku\ abundances listed in Table~\ref{tab:spec_results}.
The modeling results are shown in Table~\ref{tab:spec_evol}, and plotted
in Figure~\ref{fig:spec_evol}. We note that very similar values are found
when assuming the \chandra\ abundance set.  In addition to the \swift\ results,
Figure\,\ref{fig:spec_evol} includes the best-fit parameters for
the hotter component from the \chandra\ and \suzaku\ observations
(see subsection\,\ref{obscomp} for a discussion of similarity and differences
in the derived parameter values).

The most striking aspect of the spectral evolution is the large decrease
in the column density of the absorber, from 4.8 $\times$ 10$^{22}$ to
2.4 $\times$ 10$^{21}$ cm$^{-2}$ over the course of the observations.
This change in the absorbing column appears to be the primary reason
for the increase in flux over the first 6 weeks of the monitoring campaign.
The physical properties of the plasma component appear to be more stable
over the same time period; the temperature is approximately constant
(roughly 4 keV) between Days 61 and 125, and only begins to drop
significantly in the observation on Day 146, to 2.6 keV.  While there
is indication of a higher plasma temperature between Days 80 and 100,
the uncertainties on these temperature estimates are large, and the
values derived from the \chandra\ and \suzaku\ data during the same
time period are consistent with the 4 keV plasma being continuously
present over this time period. The emission measure (derived from the
normalization of the {\tt vapec} model and assuming a distance to the
nova of 1.4 kpc) shows some variability over the same time period.
Between Days 61 and 125, if there is a systematic decline in the
emission measure, it is masked by the scatter in our measurements.
The emission measure on Day 146 is clearly lower.

\subsection{Comparing the \swift, \chandra, and \suzaku\ results}
\label{obscomp}

The abundances found for the \chandra\ and \suzaku\ data sets, taken
at similar times, are quite different within the framework of the same
2-temperature plasma models and fixing the He, C, N, and O abundances
to values derived from optical observations.  The largest discrepancy 
in the estimate of the Neon abundance, which is a factor of 10 larger
for the \chandra\ best-fit model than for \suzaku.  The other
free-to-vary-elements differ by a factor of a few between the two models,
with the \chandra\ model having the larger abundances.  A high neon
abundance (Ne/Ne$_{\odot}$ = 95) was also reported by \citet{Tarasova14}
from optical data, although they also reported a higher iron abundance
(Fe/Fe$_{\odot}$ = 1.5) which is only consistent with our estimate
from the \chandra\ data at the 90\% upper limit.  The Fe abundance
is even lower in the \suzaku\ data. We discuss potential causes for
discrepant abundance measurements in subsection\,\ref{abundances}.
We note, however, that the absolute abundances are estimated from
X-ray data assuming hydrogen ions dominate the Bremsstrahlung continuum.
Since all ions contribute to the Bremsstrahlung continuum, and nova
ejecta have non-solar abundances, this leads to considerable systematic
uncertainties, particularly due to the non-solar abundance of helium,
which X-ray data cannot constrain. Moreover, even the determination
of the relative abundances of medium-Z elements can be challenging
if multiple plasma temperatures are present. What we can conclude
for certain is that the X-ray emitting gas is enhanced in elements
typically observed in nova ejecta.

\begin{figure*}
\centering
\includegraphics[width=4.5in]{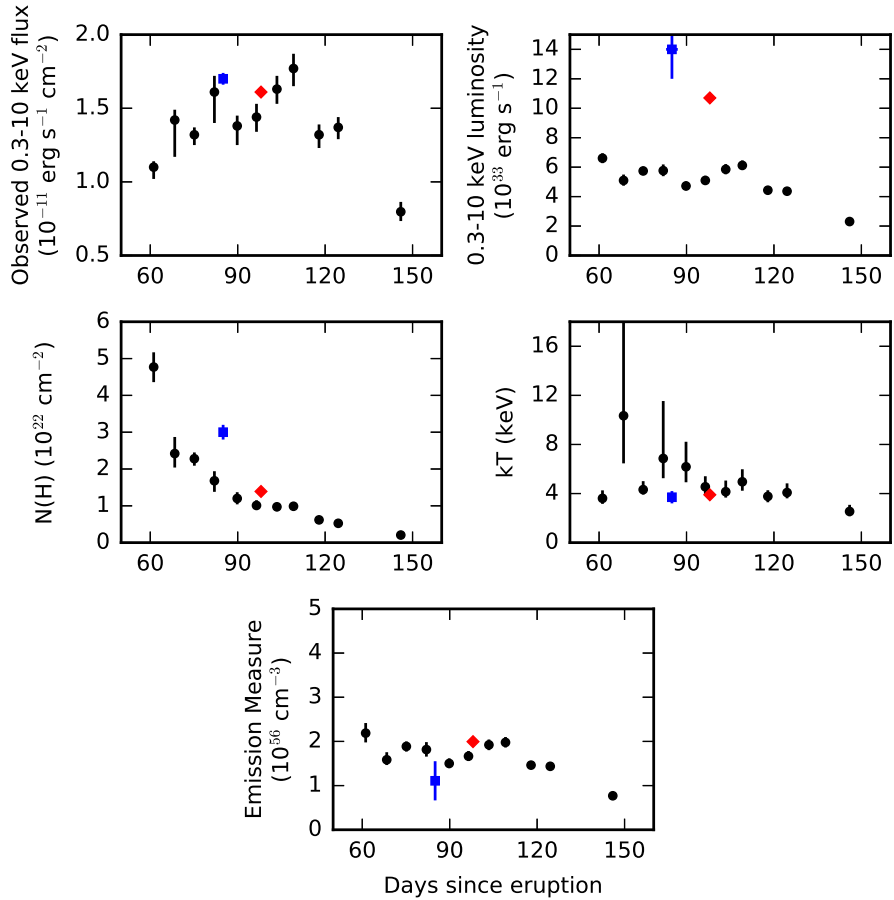}
\caption{Parameters for the hotter X-ray component derived from spectral
fits to \swift\ (black), \chandra\ (blue) and \suzaku\ (red) observations
taken between 60 and 150 days after the June discovery of the nova with
\fermi.
{\it Upper left:} The observed (absorbed) X-ray flux of the source,
measured in the 0.3--10 keV energy range. {\it Upper right} The intrinsic
luminosity of the source, determined by correcting for absorption,
for an assumed distance of 1.4 kpc.
{\it Middle left} Evolution of the absorbing column density, assuming
the ISM abundances.
{\it Middle right:} Evolution of the plasma temperature.
{\it Lower center} The X-ray emission measure, derived from
the normalization of the {\tt apec} model.}
\label{fig:spec_evol}
\end{figure*}

The agreement between the best-fit parameters obtained for the
\chandra\ and \suzaku\ spectra and those found for the \swift\ observations
taken most closely in time is good for some parameters, and
less so for others.  The \suzaku\ temperature estimate agrees closely
with that found for the \swift\ observations taken immediately before
and after, with most values falling around 3--4 keV.  The \chandra\ value
is slightly lower than its neighboring XRT values, although we note the
large uncertainties in these parameters, and the closer agreement between
\chandra\ and deeper \swift\ observations (i.e., those with smaller error
bars in Figure\,\ref{fig:spec_evol}). The \nh\ value found in the
\chandra\ observation is much higher than the \swift\ observations
taken immediately before and after.  The agreement between the
\suzaku\ and \swift\ \nh\ values is better, but still not
perfect.  Unabsorbed fluxes, and therefore luminosities are higher
in the deep spectra, but this is not surprising given the different
\nh\ values found and the fact that these models include a lower kT
component which adds additional flux at lower energies. 

The high \nh\ value found for the \chandra\ spectrum was also noted
by \citet{Peretz16}. HETG does not have sufficient sensitivity to
detect and characterize the continuum longward of 10 \AA, while it
does detect Ne lines in this spectral region. The \chandra\ spectral fits
(both our own and that of \citealt{Peretz16}) found solutions with
extremely high Ne abundances, hence very high intrinsic fluxes for the
Ne lines. The high \nh\ value is necessary to bring down the
predicted line fluxes to the observed levels.
Both \suzaku\ and \swift\ have
higher effective areas and are more sensitive to the continuum in
this spectral region. The disagreement between \chandra\ and the other
two instruments may indicate that our chosen model does not describe
the emission line ratios accurately. In terms of \nh, we proceed
assuming that the values derived from the continuum using \swift\ and
\suzaku\ data are more reliable, which also implies that the lower
Ne abundance determined with \suzaku\ is closer to the truth.

\begin{table*}
\centering
\caption{\swift\ spectral fitting results using best-fit \suzaku\ abundances}
\begin{tabular}{cccccccc}
\hline
Time & Obs ID & rate & \nh & kT & normalization$^1$ & $F_{0.3-10~ {\rm keV}}$ & $L_{0.3-10~{\rm keV}}$\\
(Day) &  & (cts s$^{-1}$) & (10$^{22}$ cm$^{-2}$) & (keV) & (10$^{-3}$) & (10$^{-11}$ \epsc) & (10$^{33}$ \eps)\\
\hline
61.2 & 00032529001 & 0.13 & 4.8 $\pm$ 0.4 & 3.6$^{+0.7}_{-0.5}$ & 0.010 $\pm$ 0.001 & 1.10$^{+0.04}_{-0.08}$ & 6.6 $\pm$ 0.3 \\
68.4 & 00032529002 & 0.17 & 2.4 $\pm$ 0.4 & 10.3$^{+10.4}_{-3.9}$ & 0.007$^{+0.0007}_{-0.0005}$ & 1.42$^{+0.07}_{-0.25}$ & 5.1 $\pm$ 0.4 \\ 
75.1 & 00032529003 & 0.17 & 2.3 $\pm$ 0.2 & 4.3$^{+0.7}_{-0.4}$ & 0.009 $\pm$ 0.0005 & 1.32$^{+0.09}_{-0.07}$ & 5.8 $\pm$ 0.3 \\
82.0 & 00032529004 & 0.24 & 1.7 $\pm$ 0.3 & 6.9$^{+4.7}_{-1.6}$ & 0.008 $\pm$ 0.0008 & 1.61$^{+0.24}_{-0.21}$ & 5.8$^{+0.4}_{-0.3}$ \\ 
89.8 & 00032529005 & 0.23 & 1.2 $\pm$ 0.2 & 6.2$^{+2.0}_{-1.3}$ & 0.007 $\pm$ 0.0005 & 1.38$^{+0.12}_{-0.13}$ & 4.7 $\pm$ 0.3 \\
96.4 & 00032529006 & 0.27 & 1.0 $\pm$ 0.1 & 4.6$^{+0.9}_{-0.7}$ & 0.008 $\pm$ 0.0005 & 1.44$^{+0.11}_{-0.10}$ & 5.1 $\pm$ 0.3 \\
103.5 & 00032529007 & 0.31 & 1.0 $\pm$ 0.1 & 4.2$^{+0.9}_{-0.5}$ & 0.009 $\pm$ 0.0005 & 1.63$^{+0.12}_{-0.10}$ & 5.8$^{+0.3}_{-0.4}$ \\ 
109.2 & 00032529008 & 0.34 & 1.0 $\pm$ 0.1 & 5.0$^{+1.0}_{-0.7}$ & 0.009 $\pm$ 0.0006 & 1.77 $\pm$ 0.12 & 6.1 $\pm$ 0.3 \\
117.9 & 00032529009 & 0.31 & 0.6 $\pm$ 0.1 & 3.8 $\pm$ 0.5 & 0.007 $\pm$ 0.0005 & 1.32 $\pm$ 0.08 & 4.4$^{+0.2}_{-0.3}$ \\ 
124.5 & 00032529010 & 0.31 & 0.5 $\pm$ 0.1 & 4.1$^{+0.8}_{-0.5}$ & 0.006 $\pm$ 0.0004 & 1.37 $\pm$ 0.08 & 4.3 $\pm$ 0.2 \\
146.0 & 00032529011 & 0.26 & 0.2 $\pm$ 0.1 & 2.5$^{+0.5}_{-0.4}$ & 0.003 $\pm$ 0.0003 & 0.80 $\pm$ 0.06 & 2.2 $\pm$ 0.2 \\
\hline
\multicolumn{7}{l}{$^1$The normalization of the {\tt bvapec} model is defined as 10$^{-14}$/4$\pi$d$^{2}$ $\int n_{e} n_{H} dV$, where $d$ is the distance}\\ 
\multicolumn{7}{l}{to the source in cm, and $n_{e}$ and $n_{H}$ are the densities of elections and hydrogen, respectively, in the}\\
\multicolumn{7}{l}{shocked plasma. All quoted uncertainties are 1$\sigma$.}
\end{tabular}
\label{tab:spec_evol}
\end{table*} 

\subsection{Constraints on the density of the nova ejecta}

\citet{Peretz16} claim that the X-ray emitting region observed with
\chandra\ in \novamon\ must have very high density of at least
6 $\times$10$^{10}$ cm$^{-3}$ (1$\sigma$; with a 3$\sigma$ lower limit of
order 10$^9$ cm$^{-3}$).  The authors derive this value
from their estimates of the Mg XI line fluxes via the He-like density
diagnostic discussed in \citet{Porquet01}, under the assumption that
the X-ray emitting region is illuminated very weakly by radiation from
the central source.  The ratio $R = f/i$, where $i$ and $f$ are the
intercombination (the middle component of the triplet) and forbidden
(longest wavelength component) emission lines of a He-like ion, is
sensitive to density under certain physical conditions.
\citet{Peretz16} find a value of $R$ = 1.40 $\pm$ 0.52 (1-$\sigma$ error),
implying the above density limit\footnote{Using the same measurement of
$R$ and the curve for T$_{\rm rad}$=0.0 and W=0.01 of \citet{Porquet01},
we derive a 1$\sigma$ density lower limit of $> 10^{12}$ cm$^{-3}$.},
and deduce that the X-ray emitting material must be highly clumped in
order to explain the observed emission measure.

Given the limited statistical quality of the HETG spectra, it is quite
likely that there is an additional uncertainty on the $R$ measurement
that \citet{Peretz16} did not account for, e.g., that due to the line
widths and shifts. To help assess the range of uncertainties on $R$, we
carried out our own fits to the Mg XI triplet. We focus only on the MEG
spectrum, binned by a factor of 2 in channel space, since there is very
little signal in the HEG spectrum at these wavelengths.  We modeled the
region of the spectrum between 8.7 and 9.45\AA\ as the sum of a power
law continuum with spectral index $-$2.5
and three Gaussian lines to represent the He-like triplet.
The rest-frame energy of all
three lines was fixed to their values as reported by the ATOMDB
website\footnote{http://www.atomdb.org/Webguide/webguide.php},
and all three components were tied to have the same blueshift,
which varied freely in the fit.
Finally, the line width was constrained
to the 90\% confidence interval found in our global model,
or 515$^{+87}_{-77}$ \kms.  In order to estimate the uncertainty
on the ratio $R$, we made use of the Markov Chain Monte Carlo
functionality in xspec, producing 20,000 realizations of the model
compared to the data and using this to determine the distribution
of $R$ that is compatible with the data.

\begin{figure}
\centering
\includegraphics[width=3in]{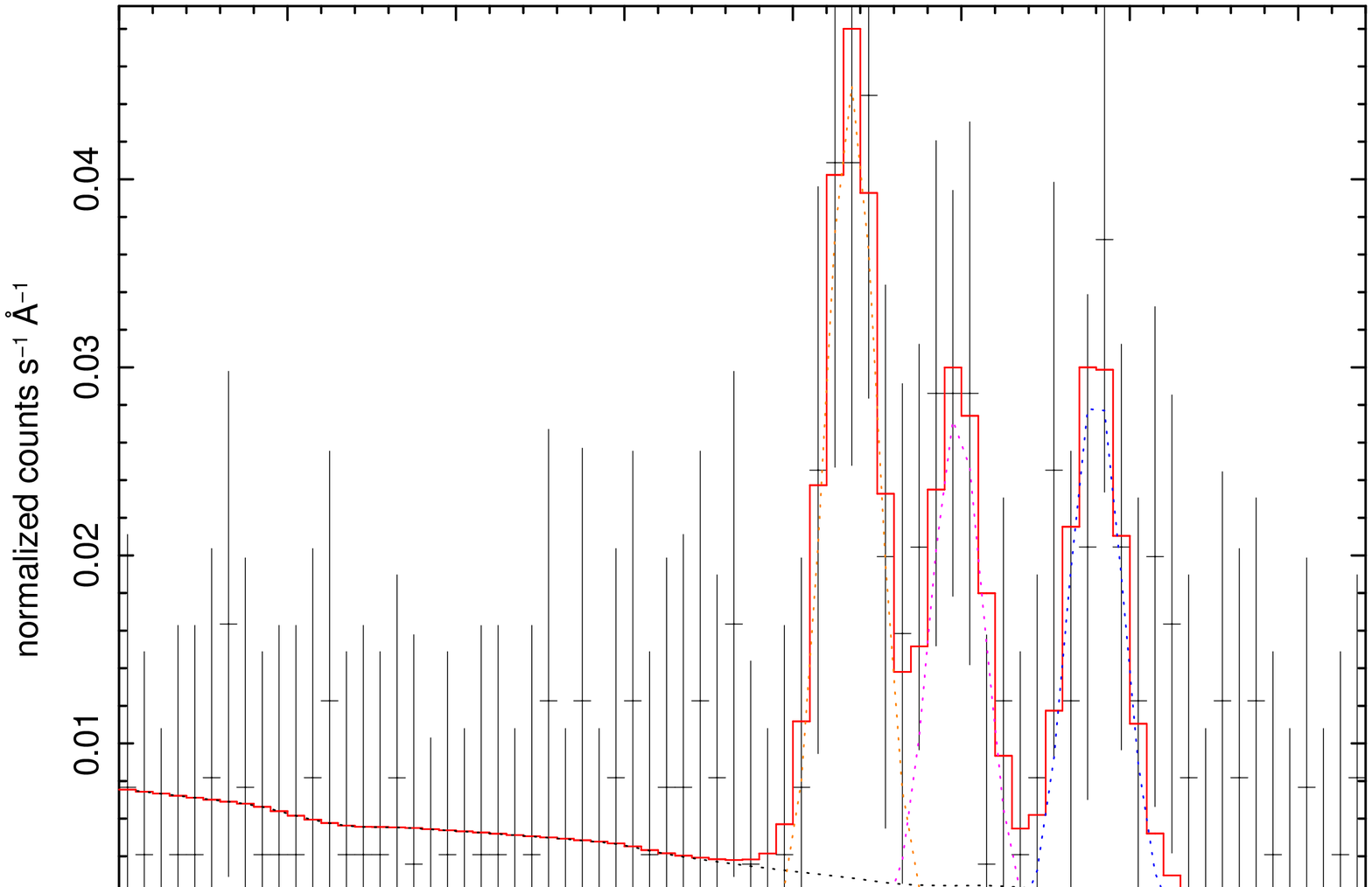}
\includegraphics[width=3in]{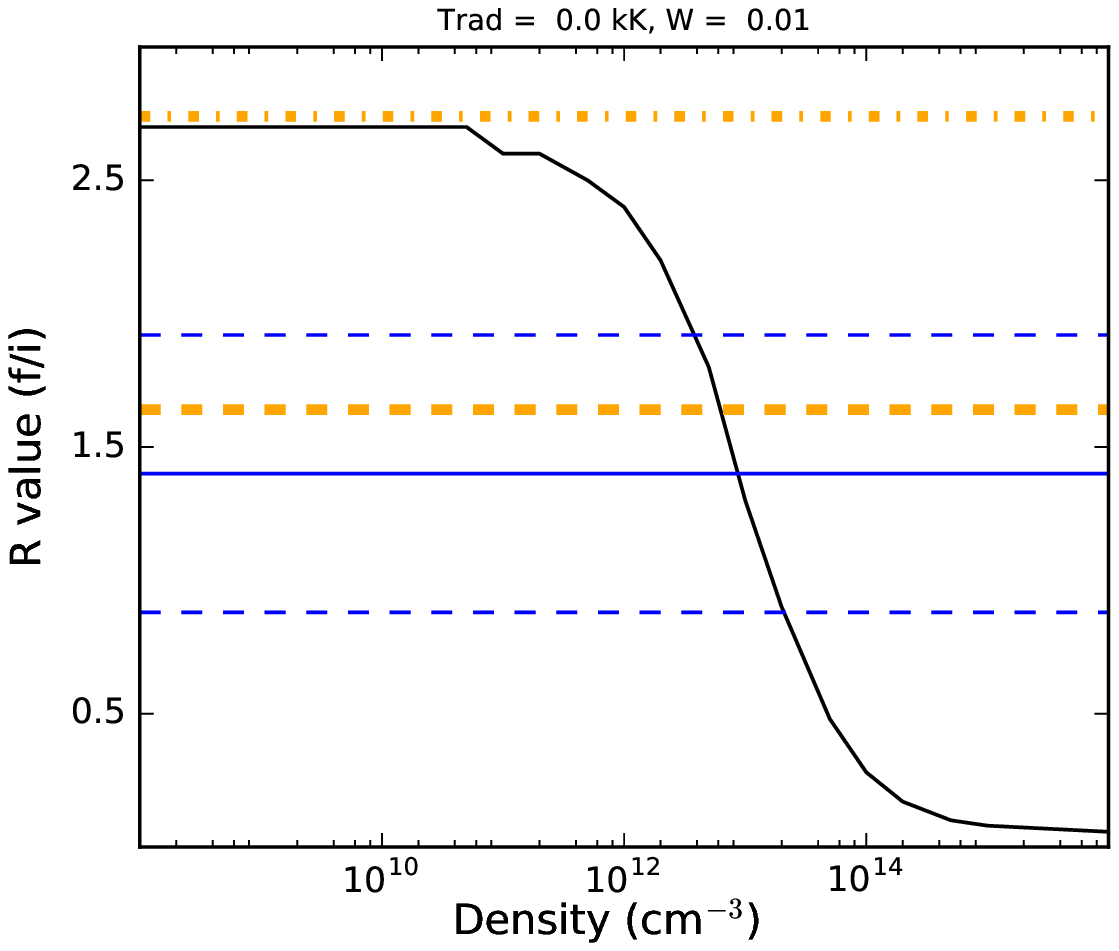}
\caption{Upper panel: Best-fit three Gaussian plus power-law model to the Mg IX
triplet region of the MEG data.
The lines (with laboratory wavelengths of 9.17, 9.23, and 9.31 \AA\
for the resonant, intercombination, and forbidden components, respectively)
are blueshifted by 1100 \kms.  Lower panel: Density diagnostic data
from \citet{Porquet01}, assuming photoionization is negligible
(T$_{\rm rad}$=0.0, W=0.01). The derived $R$-value and 1-$\sigma$
uncertainties from \citet{Peretz16} are shown in
blue as solid and dashed lines, respectively. The 1- and 2-$\sigma$
upper limits found by us are shown in orange.}
\label{fig:mg}
\end{figure}

The best-fit model is shown in the upper panel of Figure~\ref{fig:mg}.
The best-fit blueshift is larger than the value found for the global model,
with $v$ = 1100 $\pm$ 200 \kms.  We find only weak upper limits to $R= f/i$.
The 1- and 2-$\sigma$ upper limits are shown as
orange dashed and dash-dotted lines in the lower panel of Figure~\ref{fig:mg}.
At the 2-$\sigma$ level, we find that the observed lines are un-constraining
as a density diagnostic. Given the uncertainties introduced by the presence
of an unquantified radiation field, density variations in the emitting
regions, and the fact that we are approximating lines that are most
likely asymmetric with Gaussians, the constraints on $R$ are probably
even weaker than the ones we find here; ultimately, we are fundamentally
limited by the low S/N of the data. Even so, we comment on the possible
high density of the X-ray emitting region in subsection\,\ref{density},
in light of our other findings.

\section{Details of the mass ejection process in \novamon}


As we discuss in the introduction, \citet{Chomiuk14a} explained
the observed evolution of the resolved radio images of \novamon\ obtained
with the VLA with a two phase mass ejection, in which slower
material expelled preferentially in the orbital plane at early times
is followed by a period of fast mass loss.  This fast outflow leaves
the system preferentially in the polar direction, which is the path
of least resistance away from the dense equatorial material. This
dense equatorial waist (or ``torus'') plus bipolar lobe geometry was
confirmed by Hubble Space Telescope (HST) narrow-band imaging and Space
Telescope Imaging Spectrograph (STIS) spectroscopy \citep{Sokoloski17}.
Synchrotron blobs were observed at the interaction region between
these two systems of ejecta, suggesting the presence of relativistic
material accelerated in shocks at the contact surface.

We propose that the same geometry can explain the evolution we see in
the X-ray data. The shock interaction of the two systems of ejecta that
produces synchrotron emission also leads to the emission of X-rays from
the forward and reverse shock regions.  Schematically, the nova
ejecta will exist in the form of the unshocked wind, the reverse shock
front, shocked wind matter, the contact discontinuity, the shocked torus
matter, the forward shock front, and the unshocked torus matter, from
the central binary out to interstellar space.  We assume no mixing
between the torus matter and fast wind on the time scale of our
observations. This outermost layer, the unshocked warm matter
in the edge-on torus then absorbs this X-ray emission, producing
the visible spectral signature of high \nh\ material we see in
the \swift\ data.  As the equatorial torus expands away from
the central binary, the column density towards the X-ray emission drops.
In the rest of this paper, we will distinguish between the \nh\ value
of the interstellar medium (\nhism) and that of the unshocked nova ejecta
(\nhint; for ``internal'' column). We continue to use the symbol \nh\ for
the observed column, which is the sum of \nhism\ and \nhint.

The X-ray observations analyzed here provide a wealth of information
regarding the shock interaction between these two systems of outflows.
However, the situation is complex with many unknowns, including the
evolution of mass loss rate in the fast flow, so that we cannot arrive
at a unique model. We therefore focus on
aspects that can be solved with existing data with a minimum set of
assumptions. In the following, we argue that the long-lived nature of
the hot ($\sim$4 keV) component favours its origin in the reverse
shock driven into the fast outflow. We then discuss the evolution
of the absorbing column, and conclude that the slow torus was ejected
weeks after the thermo-nuclear runaway.  Furthermore, we estimate the
total mass of the slow ejecta, and also its average density at the
inner edge.  We then use the observed X-ray luminosity to constrain
the mass loss rate of the fast outflow, and its density. We further
consider the kinematics and the densities of the X-ray emitting regions.
We then briefly touch on the issue of abundances.
Finally, we discuss the
implications of our findings on the gamma-ray emission from \novamon\ in
particular, and novae in general.

\subsection{The Reverse Shock as the Likely Origin of X-rays}

In principle, the observed X-rays can be dominated by the forward
shock driven into the torus or the reverse shock driven into the
bipolar wind, or the X-rays may be due to a combination of both.
Here we propose that the predominant source of observed X-ray
emission was the reverse shock driven into the fast outflow,
based on a two stage argument. First, we argue that the long-lived
X-ray emission likely requires a long-lasting shock interaction.
We then argue that the slow torus is not physically extensive
enough to allow a forward shock to persist long enough to explain
the evolution of the X-rays. The reverse shock, on the other hand,
can exist as long as the fast wind persisted.

There was a long plateau phase during which we do not observe an
obvious downward trend in kT or luminosity, from the start of our
\swift\ observations around Day 60 to about Day 120. The plateau
phase could have started earlier in reality, but we simply do not
have any X-ray data on \novamon\ before Day 61.
If the shock had a high density
as \citet{Peretz16} inferred, the cooling time
is a fraction of a day (see subsection\,\ref{density}),
so such individual clumps would radiatively cool and cannot persist
at the same temperature for over 60 days.
If the true density of the shock was much lower, as is allowed by
our analysis, then the radiative cooling time can be
longer. However, the radius of the nova ejecta ($r$)
will have expanded by a factor of 2 (for an instantaneous
ejection at the time of the thermonuclear runaway) or more
(for a delayed ejection; see subsection \ref{delayedejection})
from Day 60 to 120; we expect this to lead to significant adiabatic
cooling. A continuing supply of freshly shocked material, heat, and
pressure is essential to maintain a roughly constant kT and a roughly
constant luminosity.

The near constancy of kT during the plateau phase suggests that the
shock velocity (the velocity of the shock front relative to the
unshocked matter) was also nearly constant during this period.
The Rankine-Hugenot conditions for a strong shock relates the
maximum shocked plasma temperature to the shock velocity as
$$v_s = \sqrt{{16\ {\rm kT}_{\rm br}} \over {3\ \mu\, m_{H}}} =  1000 \left({{\rm kT} \over {1.2 {\rm keV}}}\right)^{1/2} \left({\mu \over 0.62}\right)^{-1/2} {\rm km}\,{\rm s}^{-1} $$
where $\mu$ is the mean molecular weight of the gas and $m_{H}$ is the
mass of a hydrogen atom. For solar abundances, $\mu \sim 0.62$, which
we use here as the fiducial value, while the overabundance of metals
suggests a somewhat higher $\mu$, perhaps as high as 1.0.
The above formula incorporates the fact that the post-shock
plasma is moving at 1/4$v_s$. During the plateau phase (Day 60--120)
when the observed temperature remained kT $\sim$ 4.0 keV, the
shock velocity was 1820 \kms\ in the solar abundance case, and as low
as $\sim$1300 \kms\ if $\mu$ was close to 1.0. We must add the
velocity of the unshocked torus to estimate the velocity of the shock
front relative to the central binary, which is greater than 2000 \kms.
Thus, if the X-rays originated in the forward shock, the shock
front would have traveled about 9$\times 10^{9}$ (7$\times 10^{9}$)
km or $\sim$60 ($\sim$45) AU relative to the torus matter during
the $\sim$60 day plateau phase assuming $\mu \sim 0.62$ ($\mu \sim 1.0$).

This is larger than to the expected physical size of the torus.
If the torus was build up from $t_0$ by a constant velocity flow,
that velocity would have to have been equal to or greater than the
shock velocity to have build up a 60 AU thick torus by Day 60,
whereas a variety of clues suggests the predominant velocity of
the slow torus was less than 1000 \kms. In this case, the
outward movement of the torus during the plateau phase is irrelevant.
If, on the other hand, the torus has a range of velocities (such as
the Hubble flow type structure we adopt below), this would allow the
range of radius that the torus occupies to grow with time. However,
in this case, we expect the shock velocity to decrease as the shock front
catches up with faster and faster parts of the torus. Therefore,
the constant shock temperature is hard to explain. In either
case, the forward shock interpretation seems hard to sustain.

It is common for optical spectra of novae near maximum to show
variable outflow velocities, or of appearance and disappearance of
multiple outflow components (see, e.g., \citealt{Aydi20}).
This might bring to mind the possibility of multiple shocks.
However, the passage of a single shock front would have left the
torus in an altered dynamical state, and it seems problematic to
postulate multiple shocks through the torus with an identical
shock temperature. Moreover, optical spectra of \novamon\ throughout
the plateau phase show emission line profiles that changed little
\citep{Ribeiro13,Shore13}.

The reverse shock driven into the fast, inner flow, on the other
hand, could have moved outward or inward relative to the
central binary during the plateau phase, depending on the wind
velocity. In either case, it can maintain a constant shock
temperature, as long as the inner flow with a velocity greater
than the shock velocity persisted during the plateau phase.
Thus, we adopt the reverse shock as the likely origin of the
predominant (kT$\sim$4 keV) component of X-ray emission.
In this case, the forward shock may well be the site
of the lower temperature component identified with \chandra\ and \suzaku.
The lower shock velocity allows the forward shock to persist longer,
and the Hubble flow picture even provides a potential explanation
for the apparent decrease of the temperature of the soft component.

\subsection{The \nh\ evolution is not consistent with a shell expanding since Day 0}
\label{delayedejection}

The column density of the material absorbing the X-ray emission from
an internal shock in a nova is expected to decrease with time as the
slower ejecta expand and the density drops \citep{Mukai01}. The X-ray
monitoring of \novamon\ presented here provides one of the clearest
examples of this behavior.  For a single, thin, shell, this drop
is proportional to $r^{-2}$. However, more realistically, both the
maximum absorption and the rate at which the column density drops
depend on the mass and velocity structure of the ejected shell,
and so the observed \nhint\ values and evolution with time can be
used to place some basic constraints on the properties of the ejected
shell.  In order to do this, we modeled the \nhint\ evolution observed
with \swift\ using a simple model of a spherically symmetric shell
that expands at a constant rate with time.

We assume that the shell has an $r^{-2}$ density profile between
the inner radius $r_{\rm in}$ and the outer radius $r_{\rm out}$.
We further assume that the velocity of the expanding shell is
linearly proportional to radius, between the smallest value $v_{\rm in}$
at $r_{\rm in}$ and the largest value $v_{\rm out}$ at $r_{\rm out}$,
and the time since ejection can be used to translate between the
radius and the velocity.  This is the same ``Hubble flow'' structure
commonly used to model radio emission from novae
\citep[see][for details]{Seaquist77}. A single normalization factor
related to the total ejecta mass is then required to estimate the
integrated absorbing column through such a shell, for a given set
of $v_{\rm in}$ and $v_{\rm out}$. As the shell expands, the intrinsic
column density will drop with time, and the observed \nh\ will
asymptotically approach \nhism.

A range of estimates of \nhism\ towards \novamon\ exist
in the literature.  \citet{Shore13} favour an E(B$-$V) of 0.8 mag,
implying an \nhism\ value of 5.5 $\times$ 10$^{21}$ cm$^{-2}$
using the E(B$-$V) to \nh\ conversion of \citet{Guver09}.  \citet{Munari13}
argue for a smaller E(B$-$V) of 0.38 $\pm$ 0.01, implying \nhism\ of
just 2.5 $\times$ 10$^{21}$ cm$^{-2}$.  In the following work, we
examined three values of \nhism; 2.5, 3.5, and 5.0 $\times$
10$^{21}$ cm$^{-2}$.

\begin{figure*}
\centering
\includegraphics[width=2.3in]{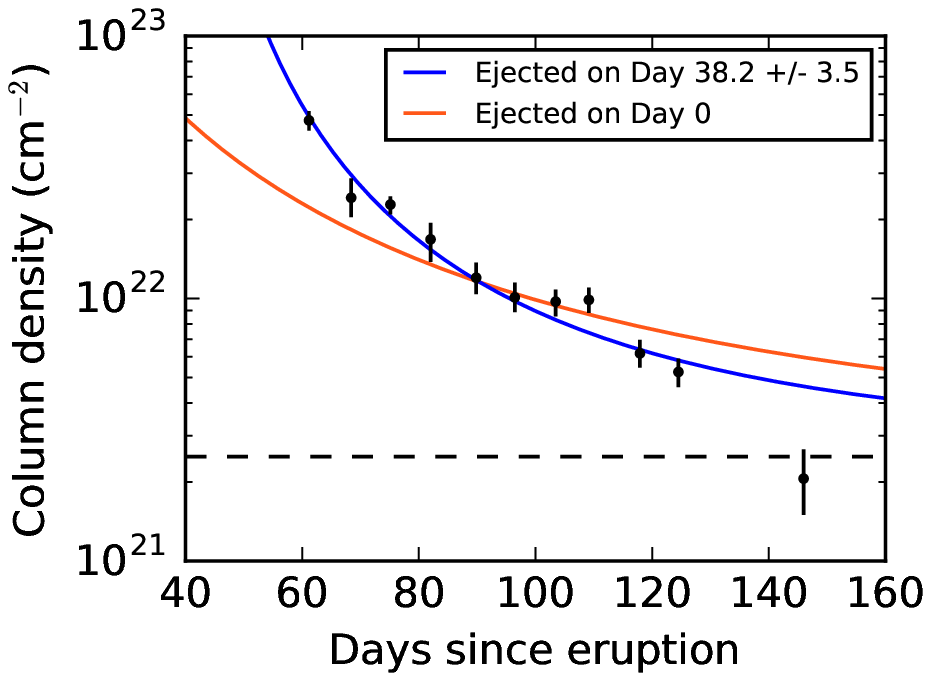}
\includegraphics[width=2.3in]{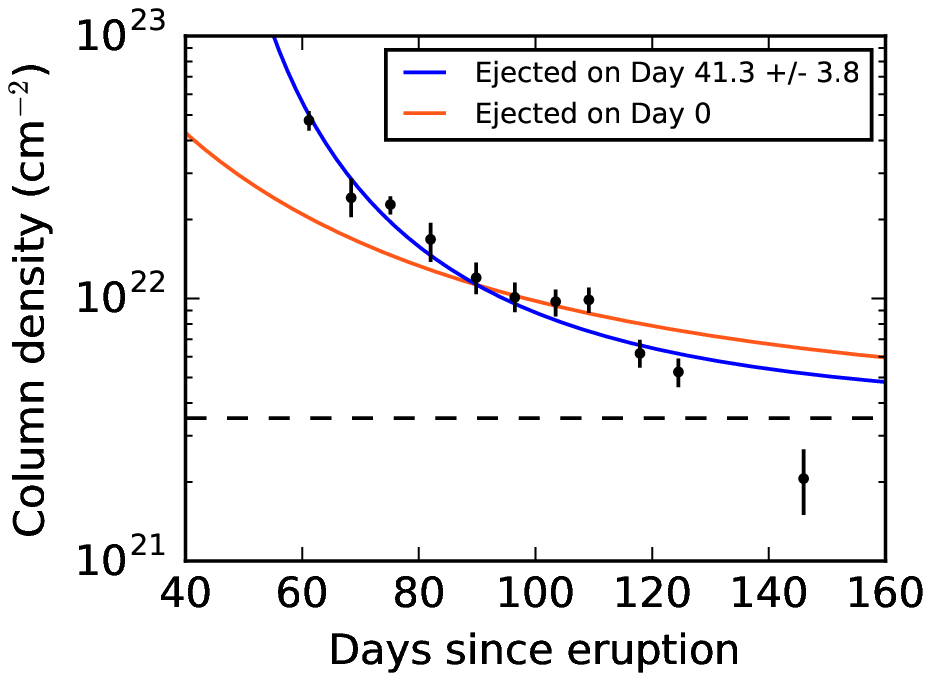}
\includegraphics[width=2.3in]{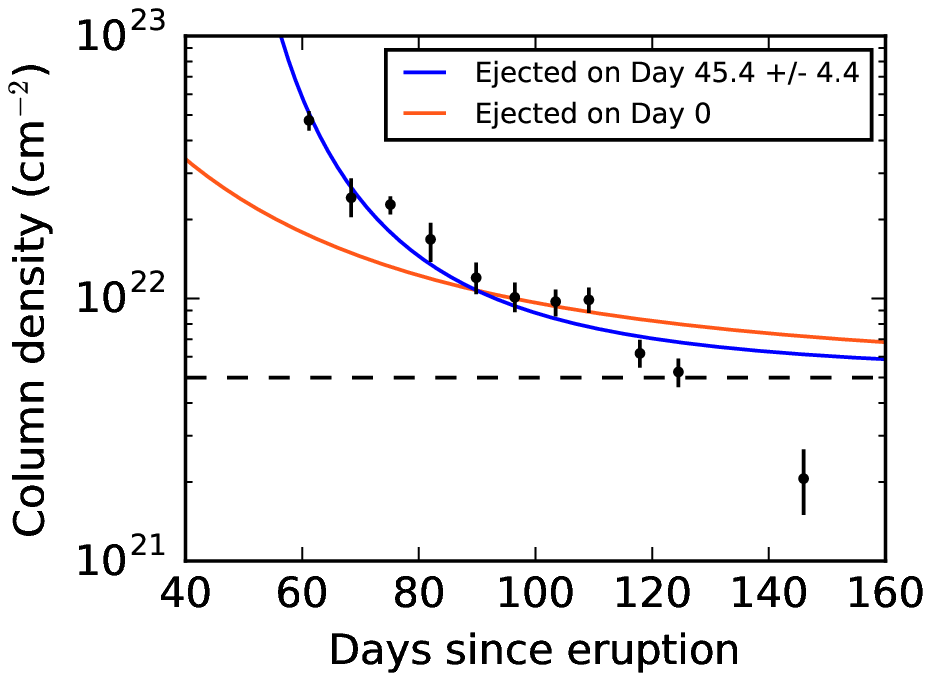}
\caption{\nh\ evolution determined from \swift\ spectra, compared with
models of \nhint\ evolution for nova shells with a $r^{-2}$ density profile
and inner (outer) velocities of 770 (2400) \kms.  We assume three values
for the interstellar column density (\nhism; shown as dashed horizontal lines).
The solid red lines show the best-fit models ejected at the time of
the \fermi\ detection, and have $\chi^{2}$/DOF values of 11.4, 14.1
and 18.8, respectively.  The blue lines are the best-fits for a delayed
expansion model, which have $\chi^{2}$/DOF values of 3.2, 4.5 and 7.5,
respectively.}
\label{fig:nh_evol}
\end{figure*}

In Figure~\ref{fig:nh_evol}, we show comparisons of the data with our
simple model of \nhint\ evolution for an $r^{-2}$ density profile shell,
assuming the three adopted values of \nhism.
Using the {\tt lmfit} package in python \citep{Newville14},
we fit the \nh\ values derived from spectral fitting of the \swift\ XRT
spectra, with the abundances found for the {\tt bvapec} model fit to
the \suzaku\ data, with our \nhint\ evolution model. We excluded the
\nh\ value obtained on Day 146 (see below). Since the ejected mass and
expansion velocity are degenerate, we initially fixed $v_{min}$ to
770 \kms\ (assuming that the blueshift of the X-ray emission is indicative
of the the inner velocity of the dense shell), and $v_{max}$ to the maximum
velocity implied by the bipolar model of \citet{Ribeiro13}, 2400 \kms.
We then fit for the ejected mass.  The best-fit \nh\ found on Day 146
lies well below the model in all cases.  This value is also lower than
the \nh\ reported for later \swift\ observations during the supersoft
phase of the nova evolution. We speculate that the \nh\ derived
for this spectrum is anomalous due to the earliest presence of the
photospheric emission, even though it is not clearly recognizable as
such, making the spectrum flatter and the implied \nh\ smaller. 

Regardless of the assumed \nhism\ value, shell models
ejected at t$_{0}$ (i.e. the time of the \fermi\ detection)
do not give a good fit to the best-fit column density values.
These models, shown as the red lines in Figure~\ref{fig:nh_evol},
decline at a slower rate than the data values, and have difficulty
accounting for the high \nh\ observed before Day 80.
Therefore, we also fit models that leave the start date
of the expansion of the ejecta (hereafter $t_e$) a free parameter
(fixing the \nhint\ to a high value before that time). The delayed
ejection models result in a much better fits to the data at early
times than the t$_{0}$ ejection models, for all assumed values of
\nhism.  The derived delay times get later as the assumed \nhism\ increases
in value, from 38 days for \nhism\ = 2.5 $\times$ 10$^{21}$ cm$^{-2}$ to
45 days for \nhism\ = 5 $\times$ 10$^{21}$ cm$^{-2}$. For each assumed
value of interstellar absorption, \nhism,  the ejection time is constrained
by the decline rate of the observed \nh\ values (as long as the shell
expands at constant velocity), and so is independent of the assumed
inner and outer velocities of the shell. The overall best-fit to the
data is obtained for \nhism\ = 2.5 $\times$ 10$^{21}$ cm$^{-2}$.

Absorption of X-rays with photon energies in the $\sim$1 to several keV
range is largely due to K shell electrons of medium-Z elements, such as
Oxygen. If, during the course of \swift\ XRT monitoring, an increasing
fraction of the slow torus becomes shocked, and therefore become ionized,
then the measured \nh\ values would decline faster than the above analysis
would suggest. This effect is unlikely to explain the discrepancy between
the measurements and the prompt ($t_e$=$t_0$) ejection model: if this was
the reason for the faster-than-expected decline of \nhint\ before Day 82, we
expect the decline to continue to be faster than the unshocked prompt
ejection model after Day 82, which is not what we observe.
On the other hand, it is possible that the ionization of the slow torus
is in part responsible for the poor agreement between the data and the
models (prompt or delayed) at late times, say after Day 110.

We note that \citet{Linford15} also found evidence for a delay in
spatial expansion of the radio-emitting nova shell of \novamon\ in
VLA images obtained during Days 126--199. During these times, the
VLA images are elongated in the east-west direction due to the spatial
extension of the fast outflow. The angular diameters in the north-south
direction of both the eastern and western lobes are seen to expand from
0.06 to 0.11 arcsec. The measurements are consistent with expansion at
constant velocity that started at Day 25 $\pm$ 10 (see Figure 8 of
\citealt{Linford15}). Since this refers to the fast outflow, while
the delay inferred from the X-ray \nh\ evolution is that of the slow
torus, they are not direct corroborations of each other. Nevertheless,
it seems encouraging that independent observations of two distinct
components of outflow in \novamon\ both indicate delayed ejection,
perhaps suggesting a common origin for the delay.

Recently, \citet{Sokolovsky20} also inferred a delayed ejection of the torus
in V906~Car (=ASASSN-18fv), by analyzing the \nh\ evolution of its X-ray
emission. Moreover, a similar delayed ejection scenario was
proposed to explain the concurrent radio and X-ray evolution of
T Pyx during its 2011 eruption \citep{Nelson14, Chomiuk14b}.
In that system, radio emission did not start rising until $\sim$60 days
after eruption, based on radio flux density evolution and the
late onset of hard X-ray emission. Stalled expansion
may be common in novae, and implies a phase where common envelope physics
can influence mass ejection and angular momentum loss from the system.

Our working model, then, is that the thermonuclear runaway can
leave some fraction of the accreted matter in an extended, quasi-static
configuration, having a red giant-like dimension. Optical spectroscopy
measures the expansion velocity of the part of the envelope that is
promptly ejected to infinity, but we do not have a direct velocity
measurement of the matter responsible for the optical continuum.
We postulate that the continuum source is the quasi-static,
marginally-bound envelope, to borrow the terminology of \citet{Pejcha16}.
Such an extended envelope can easily be ejected later, following the
injection of additional energy, such as via common-envelope interaction
or by residual nuclear burning. However, 
the delayed ejection model so far is purely phenomenological, and
we do not currently have a quantitative prescription
for the delay time.

\subsection{The mass and the density of the torus}
\label{ejectamass}

We now use the normalization of the \nhint\ evolution model to estimate
the total ejecta mass of the torus. Since X-ray absorption measures
the total gas+dust mass with only weak dependence on the amount or the
properties of dust \citep{Wilms00}, our estimate is insensitive to any
production of dust by \novamon. Since we can only observe the
\nhint\ evolution as seen from the Earth, we make the simple assumption
that the torus is a partial spherical shell that covers some unknown
fraction of the solid angle as seen from the central binary, but otherwise
uniform in all directions. For $v_{min}$ and $v_{max}$ of 770
and 2400 \kms, respectively, the best fit ejected masses range from
(3.8 $\pm$ 0.8) $\times$ 10$^{-5}$ M$_{\odot}$
(\nhism\ = 2.5 $\times$ 10$^{21}$ cm$^{-2}$)
to (1.7 $\pm$ 0.8) $\times$ 10$^{-5}$ M$_{\odot}$
(\nhism\ = 5 $\times$ 10$^{21}$ cm$^{-2}$). 
These solutions are not unique, and essentially scale with the ratio
$v_{max}/v_{min}$. Since \nhint\ scales with the total torus mass
divided by the characteristic surface area, and since lower expansion
velocities result in lower surface areas at any given time after eruption,
a lower total shell mass is required to match the observed \nhint\ values
if $v_{max}$ is lower.  Note that $v_{max}$ here refers to the maximum
velocity of the torus material, which may well be lower than 2400 \kms,
the fastest velocity inferred for this nova. This will result in
a lower estimate for the ejecta mass: Using an extreme assumption of
$v_{max}$=$v_{min}$=770 \kms, we find a minimum ejected mass of
(6 $\pm$ 2) $\times$ 10$^{-6}$ M$_{\odot}$ for the highest assumed
value of \nhism. 

This range of ejected masses (1.7 -- 3.8 $\times$ 10$^{-5}$ M$_{\odot}$)
is compatible with the ejected shell estimate of 4 $\times$ 10$^{-5}$
M$_{\odot}$ presented by \citet{Chomiuk14a}, although we note the radio
observations are sensitive to all ejected material (i.e. both the torus
and the bipolar outflow) while our \nhint\ measurements trace only the
slower, outer torus.  Further consideration, however, suggests a potential
inconsistency.

First, our model assumes a spherical shell; the total mass absorbing
the X-rays is likely lower since the torus is not filling the full
spherical volume around the central binary.  The HST imaging gives
a sense of the opening angle of the torus \citep{Sokoloski17}. The
true torus mass is smaller than what we derive assuming a sphere,
by a factor whose exact value is unknown but is probably of order
0.5--0.2.

There is another significant correction factor based on the composition of
the slow torus. As we noted above, X-ray absorption is primary due to
medium Z elements, which we (as well as \citealt{Peretz16}) find to be
over-abundant in the X-ray emitting plasma. Moreover, \citet{Sokolovsky20}
found direct evidence of non-solar abundance absorber in the
\nustar\ observations V906~Car. For \novamon,
we do not have any direct evidence that the X-ray absorber has non-solar
abundances. Because of this, and because we do not have precise and
accurate measurements of the abundances of relevant elements (e.g.,
the strong disagreements on Ne abundance between \chandra\ and
\suzaku\ data), we have retained the solar-abundance absorber as
our baseline. We did perform one experimental fit to the \suzaku\ data
by separating the absorber into the interstellar component (with
standard abundances according to \citealt{Wilms00} and \nhism\ fixed
at 2.5 $\times$ 10$^{21}$ cm$^{-2}$) and the torus, whose abundances
are tied to those of the {\tt bvapec} component. We find
approximately a factor of 5 lower \nhint, and hence the
total mass, of the torus. We take this as a representative factor in
the possible overestimation of the total torus mass estimated using
the X-ray absorption.

Combining both the geometrical and abundance factors, with the
estimate of (3.8 $\pm$ 0.8) $\times$ 10$^{-5}$ M$_{\odot}$
(\nhism\ = 2.5 $\times$ 10$^{21}$ cm$^{-2}$),
we obtain a revised estimate of the slow torus mass of
(1.5$\pm 0.2) \times 10^{-6}$ M$_{\odot}$ (for a geometrical factor
of 0.2) to (3.8$\pm 0.5) \times 10^{-6}$ M$_{\odot}$
(for 0.5). These values are notably lower than the radio estimate
of 4 $\times$ 10$^{-5}$ M$_{\odot}$ \citep{Chomiuk14a}. Even though
both X-ray and radio estimates have large error bars, this discrepancy
is significant enough to be worrying. Here we consider the possible origins
of this disagreement.

On the radio side, the sources of uncertainty include the distance $d$ to
\novamon, the range of expansion velocities, and the filling factor $ff$ of
the ejecta (see, e.g., \citealt{Nelson14} who explored the various sources
of uncertainties in the context of their study of radio emission from
T~Pyx).
A torus of mass $3 \times 10^{-6}$ M$_\odot$
with dynamical parameters we have assumed in modeling the \nhint\ evolution
would have become optically thin at 1.8 GHz by Day 100 if $ff$ is 1.0,
Day 150 or so if $ff$=0.1.  This is clearly inconsistent with the
observed late-time behavior of the multi-wavelength radio light
curve, reinforcing the severity of mass discrepancy between
X-ray and radio data.

On the X-ray side, two obvious sources of systematic uncertainties
are the expansion velocity and the $r^{-2}$ radial profile that we
assumed. For a given ejecta mass, the absorbing column density would
be lower if the typical distance from the central binary (and hence
the surface area over which the mass is distributed) is larger.
Therefore, we can raise the X-ray estimated mass and make it closer
to the radio estimated mass by assuming a larger $v_{min}$ as perhaps
suggested by HST/STIS spectroscopy \citep{Sokoloski17}. However, the
radio imaging data \citep{Linford15} constrain the ratio of $v_{min}$
to $d$ ($\sim$600 \kms\ for 1.4 kpc). If we attempt to raise the
X-ray estimate of the torus mass by assuming a much larger $v_{min}$,
$d$ must also be larger by the same factor, which then increases the
radio mass estimate, so this would not help solve the discrepancy.
Similarly, if the radial density profile was radically different from
$r^{-2}$ such that the typical distance of absorber from the central
binary was greater by a factor of 3, that would increase the X-ray
estimate of the mass by roughly the required amount. However, this
would also make the radio image more extended, unless the $d$ was greater.

We may be able to reconcile the two estimates by assuming the torus
to be clumpy.  The radio estimate of the mass scales as $ff^{0.5}$,
under the assumption that a fraction $ff$ of the volume contains
all the mass, and \citet{Chomiuk14a} assumed $ff \sim 0.1$, following
\citet{Shore13}.  If clumps were small and numerous, it is likely that
the effects of clumping would average out in terms of X-ray absorption,
and that all lines of sight would have the same integrated \nhint; this would
not change the X-ray estimate of the torus mass. However, suppose that the
number of clumps was lower such that many lines of sight to the X-ray
emitting region did not intersect any clumps, while most of the others
intersected one clump. Further suppose that \nhint\ through any single
clump was high enough (say $> 5 \times 10^{23}$ cm$^{-2}$) to absorb
all photons below $\sim$5 keV. The radio data are analyzed assuming
all the mass is contained in the clumps \citep{Nelson14}, which is
justified if the density contrast is strong enough. The X-ray analysis,
in this situation, reveals the mass contained in inter-clump medium.
If so, there is no reason to expect the radio and X-ray measurements
to agree with each other. Moreover, we may be able to explain
the high energy excess seen in the \suzaku\ data (see
section\,\ref{suzakuspec}) as due to X-rays passing through such
high density clumps.

We can also estimate the density of the inner edge of the torus
using the same model. The value of
\nhint\ is obtained by integrating the density between the inner
($r_{\rm in}$) and outer ($r_{\rm out}$) radii of the torus.
In a Hubble flow model with an $r^{-2}$ density structure, this
can be written in terms of the density of the inner edge of the
torus $n_{\rm in}$ as
\nhint\ = $n_{\rm in}$ $r_{\rm in}^2$ (1/$r_{\rm in}$ $-$ 1/$r_{\rm out}$).
For $t_e$=40 days and $v_{min}$= 770 \kms, $r_{\rm in}$ is 1.4 $\times 10^{14}$
cm and 5.6$\times 10^{14}$ cm, respectively, at Days 61.2 and 124.5,
Ignoring the 1/$r_{\rm out}$ term, the measured
\nh\ values (4.8 $\times 10^{22}$ cm$^{-2}$ and
0.5 $\times 10^{22}$ cm$^{-2}$), minus the assumed \nhism\ 
(2.5 $\times 10^{21}$ cm$^{-2}$), translates to the inner edge
densities of 3.3$\times 10^8$ and 4.5$\times 10^6$ cm$^{-3}$,
respectively, before the abundance-related correction factor.
These estimates are inversely proportional to the assumed inner velocity
of the slow torus.

\subsection{The density of the reverse shock and its implications}
\label{density}

\begin{figure}
\centering
\includegraphics[width=8.2 cm]{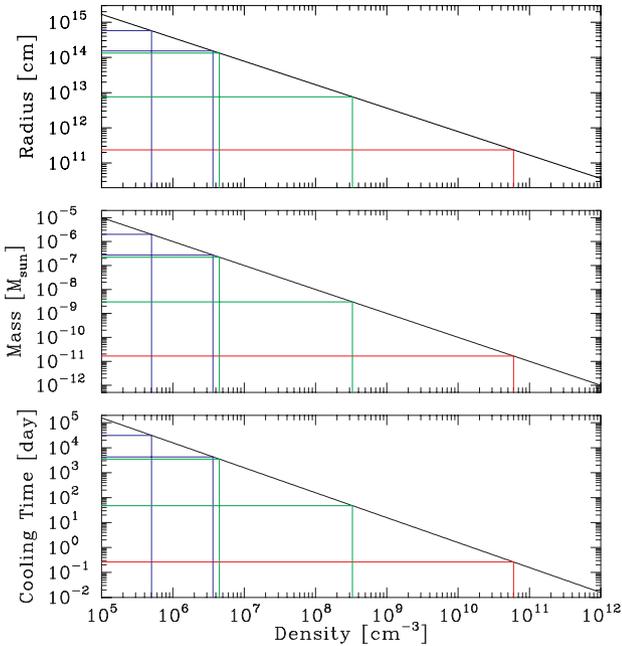}
\caption{Three properties of the post-shock plasma are shown as a function
of the assumed post-shock density, knowing that the emission measure was
roughly 2$\times 10^{56}$ cm$^{-3}$ during the plateau phase. The top
panel shows the radius of a sphere whose total volume is required to
explain the observed emission measure. The middle panel shows the total
mass of the shocked matter. The bottom panel shows the Bremsstrahlung
cooling time for a 4 keV plasma. See text for the explanations of blue,
green, and red lines}
\label{fig:xfig}
\end{figure}

In contrast to the torus matter, we do not have direct observables
that allow us to constrain the density of the reverse shock tightly.
One can, however, estimate several key properties of the shock as
a function of assumed post-shock density by noting that the observed
emission measure was roughly constant (2$\times 10^{56}$ cm$^{-3}$
for a distance of 1.4 kpc) during the plateau phase. Since the X-ray
emission measure is the density squared integrated over the emission
volume, one can relate the density and the volume by further assuming
that both the temperature and the density of the X-ray emitting plasma
was uniform. Knowing the density and the volume, one can estimate the
total mass of the shocked plasma as well, which, for a constant emission
measure (=density$^2 \times$ volume = density $\times$ mass), is inversely
proportional to density. Furthermore, one can estimate the cooling time;
here we use the Bremsstrahlung cooling time, which is a reasonable
approximation for a 4 keV plasma, and is given by 
$$ t_{\rm{cool}} = \left({n_e \over 6.8 \times 10^{14}}\right)^{-1} \left({{\rm kT} \over 1 {\rm keV}}\right)^{1/2} {\rm s} $$
We show these relationships in Figure\,\ref{fig:xfig}.

The delayed ejection models provide estimates of the the inner radius
of the absorbing matter as a function of time.  Using the version shown
in the left panel of Figure~\ref{fig:nh_evol}, the inner radius was
1.5$\times 10^{14}$ cm on Day 61.2 and 5.7$\times 10^{14}$ cm on Day
124.5, times of the two \swift\ observations that define the plateau
phase. These radii are indicated by the horizontal blue lines in the
top panel of Figure\,\ref{fig:xfig}. The vertical blue lines indicate
the density that the post-shock plasma needs to have to explain the
observed emission measure, if it occupied the entire spherical regions
inside these radii. In reality, the post-shock region occupies a fraction
of such a sphere, so the true density is to the right of these lines.
In the middle and bottom panels, blue lines indicate the total shocked
mass and cooling times that are implied by these two limits on the
density.

These lower limits on the density allow us to comment on the possibility
that the X-ray spectrum may be due to a single-temperature, non-equilibrium
ionization plasma. Since $nt \sim 10^{12}$ cm$^{-3}$s is the condition
to reach ionization equilibrium \citep{Smith10}, it takes a few days
to do so at the beginning of the plateau phase. It is therefore unlikely
that non-equilibrium effects are
important for our observations of \novamon, and justifies our choice
of 2 temperature collisional ionization equilibrium model.

\citet{Peretz16} inferred a 1$\sigma$ density lower limit of the X-ray
emitting plasma of 6$\times 10^{10}$ cm$^{-3}$ (shown in red in
Figure\,\ref{fig:xfig}), our own analysis suggests this was probably
not a secure result. Instead, we rely on the estimates for the density
of the inner edge of the torus, obtained by considering the \nhint\ evolution,
of 3.3 $\times 10^8$ and 4.5 $\times 10^6$ cm$^{-3}$ at Day 60 and 120,
respectively (shown in green).  For the reverse shock to dominate the
X-ray emission throughout the plateau phase, as we argued based on the
longevity of this phase, the density of the reverse shock must be lower
than that of the inner edge of the torus. This results in the fast flow
immediately slowing down, thus ensuring a strong shock. In contrast, the
additional momentum provided by the fast flow results only in a small
perturbation of the slow torus, so the forward shock makes only a minor
contribution to the observed X-rays.  This suggests that the density of
the reverse shock to be left of the green vertical lines.

We can now use the constraints on the post-shock densities of the
reverse shock to infer the mass-loss rate, assuming a spherical outflow
with a constant mass-loss rate. A fiducial rate of 1.0 $\times 10^{21}$
g\,s$^{-1}$ would results in post-shock densities of $\sim 4.0 \times 10^7$
and $\sim 2.5 \times 10^6$ cm$^{-3}$ at Days 60 and 120, respectively.
Comparing these estimate with the lower (blue) and upper (green) limits on
the post-shock densities of the reverse shock we estimated above,
we see that any constant mass-loss rates within a factor of 2--3 of
1.0 $\times 10^{21}$ g\,s$^{-1}$ would have densities that satisfy
both limits through days 60--120. Such an outflow implies a shock
power of order 10$^{37}$ \eps, considerably higher than the observed
X-ray luminosity, much higher than the observed luminosity of
$\sim 6 \times 10^{33}$ \eps; this is consistent with the long cooling
time inferred for the relatively low densities. For a radiative
shock to explain the observed luminosity, one needs the $\sim$1820 \kms\ matter
to be shocked at a rate of roughly 4$\times 10^{17}$ g\,s$^{-1}$.
If this was a uniform density, spherical flow, at $r_{\rm in} \sim 1.4 \times
10^{14}$ cm (appropriate for Day 60, $t_e$=40 d, $v_{\rm in}$=770 \kms),
the post-shock density is $\sim 3.6 \times 10^4$ cm$^{-3}$. This requires
clumps with densities 4 orders of magnitudes higher for such a shock to be
radiative, exceeding the estimated density of the inner edge of the torus.
This interpretation almost certainly requires small, high density clumps
in the fast outflow to collide with high density clumps in the torus,
a situation we find unlikely.

\subsection{Kinematics of the Two Outflows}

The blueshift and the broadening of the emission lines have the
potential to confirm, or refute, the scenario outlined above.
First we note that the line width is not due to thermal Doppler
motions. While the thermal velocity of hydrogen is $\sim$700 \kms
for a 4 keV plasma, the thermal velocity of an ion scales with the
square root of the atomic weight, so it is $\sim$140 \kms\ for Mg,
too small to be the origin of the measured line width. Instead, the
line width is due to the bulk motion of ions moving
in a variety of directions. The fact that we observe a net blueshift
requires that redshifted emission be hidden. In an expanding shell,
ions moving away from us are on the far side of the central binary.
Therefore, we posit that the unshocked part of the fast outflow
at the center of the nova occupies a large enough volume to
be able to absorb the redshifted emission on the far side, allowing
us to observe only the blueshifted emission on the near side.

If the bulk of the X-rays are emitted by the forward shock driven into
the torus, and if the unshocked torus matter is moving outward at 770 \kms,
then the shock front is traveling outward at an inertial frame velocity of
$\sim$2600 \kms\ (for solar abundances) or $\sim$2100 \kms\ (for highly
metal enhanced case), given kT$\sim$4 keV. The strong shock condition then
means that the post-shock, X-ray emitting plasma is moving outward at
a velocity of 2150--1750 \kms.
This does not necessarily imply
that we should observe a net blueshift of $\sim$2000 \kms:
What we observe is the average of projected velocities of all observable
parts of the torus, including the part that is
moving away from us. The net blueshift can be reduced if more of the
torus is observable, while the line width can be reduced if only a small
portion is observable: it does not appear feasible to reduce both by
absorption.

To assess the expected kinematics of the reverse shock matter, we rely
on the measurements of ejecta velocities in the system from optical
spectroscopy (maximum expansion velocity of 2400$^{+300}_{-200}$
\kms\ according to \citealt{Ribeiro13}).
If the faster ejecta have this maximum velocity, and the shock velocity
is 1820 \kms, and if the reverse shock is moving with the inner edge of
the slow torus, then the slower material being swept up must be traveling
at $\sim$580$^{+300}_{-200}$ \kms. The observed lines would include
contributions from the lower-temperature matter in the forward shock,
whose bulk outward velocity should be of order $\sim$1000 \kms\ using
a shock temperature of order 0.5 keV and the same argument we used
in the previous paragraph. The observed kinematics of the emission
regions does not appear to contradict our reverse-shock dominated
model, within the limit of the \chandra\ HETG data.
Higher quality X-ray spectroscopy of a future bright nova has the
potential to distinguish these two scenarios unambiguously.

\subsection{Abundances of the ejecta}
\label{abundances}

As noted in section\,\ref{xanalysis}, the abundances we derived
using \chandra\ and \suzaku\ data do not agree quantitatively.
Moreover, while optical spectra \citep{Tarasova14} suggest a high
abundance of iron, \suzaku\ data in particular indicate a lower
than solar abundance of iron. Similar tension regarding iron
abundance between the optical and X-ray data were noted for
V382~Vel \citep{Mukai01}. Different sensitivity of \chandra\ and
\suzaku\ data to different features (continuum vs. lines,
hard energy bands vs. soft) and different assumptions made during
data analysis may in part be responsible. As we argued earlier,
this is likely to be the case for the disagreement regarding the
abundance of neon, coupled with the disagreement regarding \nh,
between \chandra\ and \suzaku\ data. However, the X-ray vs. optical
disagreements regarding iron abundance, in particular, has now been
found in multiple novae. For V382~Vel, \citet{Mukai01} proposed
a possible ionization effect, but it may also exist in the X-ray
absorber in V906~Car \citep{Sokolovsky20}.

The disagreements may be due to the chemical inhomogeneity of nova ejecta,
created through gravitational settling, diffusion and/or mixing of core
material prior to and during thermo-nuclear runaway, and the nuclear
reactions during runaway itself.
Perhaps abundance differences among different layers can
persist in the nova ejecta.
Abundance measurements of spatially
resolved nova shells would be a great first step in assessing if
this is a real effect.

\subsection{Implications for gamma-ray emission phase of \novamon\ and other novae}

The luminosity of the shock X-ray emission we observe in \novamon\ is below
10$^{34}$ \eps\ ever since the \swift\ monitoring started. This is similar
to those of shock X-rays observed in other novae with dwarf mass donors
(i.e., cataclysmic variables; those in symbiotic systems, such as RS Oph,
are more luminous; see \citealt{Mukai08}) at similar stages of nova eruptions.
On the other hand, we have inferred that the X-rays we observe in \novamon\ 
after Day 60 were probably from a non-radiative shock, and the shock power
can be as high as 10$^{37}$ \eps. While this approaches the shock power
needed to explain the gamma-ray luminosity, one would not expect the
shock to have been radiatively as inefficient at early times. This
is because gamma-ray emission epoch is before the ejection of the torus
from the binary, meaning that the shock is much closer to the white dwarf.
Therefore, the post-shock density would have been higher, and hence a much
higher fraction of the shock power should be emitted promptly as X-rays. 

We now have two \nustar\ detections of novae concurrent with \fermi\ detection
of GeV gamma-rays.  The thermal X-ray luminosity of V5855~Sgr was
found to be a few times 10$^{33}$--10$^{34}$ \eps\ for a distance of 4.5 kpc
in a single component fit to the \nustar\ spectrum \citep{Nelson19}. The X-ray
luminosity of V906~Car was found to be $\sim$1.5 $\times 10^{34}$ \eps\ for
a distance of 4 kpc in a single-temperature fit \citep{Sokolovsky20}.
These luminosity values are similar to what we infer for \novamon\ during
the observations described here. Moreover, \citet{Sokolovsky20} found
a smooth evolution of X-ray properties from the early times, when
the spectrum was so highly absorbed as to be detectable only with
\nustar, to the later epochs when less absorbed X-rays were detected
with \swift.  This suggest that the early X-rays, observed concurrently
with GeV gamma-ray emission, also originate in the reverse shock.

While the true thermal X-ray luminosity could have been higher in both
V5855~Sgr and V906~Car at the time of \fermi\ detection,  if
more complex spectral models were considered, such models were not required
by the \nustar\ data. In addition, to our knowledge, the only confirmed
nova ever detected with all-sky X-ray monitors is RS~Oph (with \swift\ BAT;
\citealt{Bode06}) even though these all-sky surveys
are eminently capable of detecting transient X-ray sources with luminosities
of order 10$^{36}$ \eps\ at Galactic center type distances, unless
X-ray transients CI Cam and MAXI J0158$-$744 turn out to be
nova eruptions as interpreted by several authors
\citep{Ishida04, Li12}. These
observations are placing an increasingly tighter constraints on the
regions of phase space where unseen powerful ($\sim 10^{38}$ \eps)
shocks can exist in novae during the gamma-ray emission phase.
Optically-thin, thermal X-rays from such shocks can only hide if the
temperature was lower than that of the X-rays we do detect later in
the life of a nova, so that it can be easily absorbed by a modest
absorption column, or if their luminosity decayed rapidly
while the absorbing column through the torus is still high.

\section{Conclusions}

We have analyzed X-ray observations of the \fermi-detected nova
\novamon\ obtained with \swift, \chandra, and \suzaku\ between 60 and
160 days after the start of the eruption.  Our major conclusions are as follows:

\begin{itemize}
\item The early X-ray evolution observed in \novamon\ is consistent with
internal shocks in the ejecta as the fast outflow collides with
the slower material in an equatorial torus.  This picture is broadly
consistent with the what is already known about the nova from radio and
optical observations.

\item We require a multi-temperature model to explain the line strengths
of H- and He-like ions of Mg and other medium-Z elements. We propose that
the hotter component, responsible for most of the flux, is from the reverse
shock driven into the fast outflow, while the lower temperature component
is due to the forward shock driven into the torus.

\item The absorption towards the X-ray emitting region drops with time
as the ejecta expand and decrease in density. The steep decline of
\nhint\ observed in the \swift\ data is not consistent with a shell that
expands from $t_{0}$; instead, the decline is better matched by a shell
that was ejected from the white dwarf roughly 38 days after the start
of the eruption.

\item The observed X-ray spectra are consistent with the overabundance
of medium-Z elements. If the X-ray absorber has the same non-solar
abundances, and allowing for the geometrical factor of the torus,
the mass of the shell ahead of the X-ray emitting region is perhaps
a few times 10$^{-6}$ M$_{\odot}$, considerably lower than the radio
estimate of the total ejecta mass. We consider the possibility that
the torus is clumpy: the radio emission was dominated by the high-density
clumps, while the X-ray absorption was dominated by the inter-clump
material.

\item At the time of the \swift\ observations, the X-ray emitting
shock probably was not radiative.  Although the observed X-ray luminosity
was modest (below 10$^{34}$ \eps), the true shock power was likely
considerably higher, perhaps 10$^{37}$ \eps. The inferred shock power
is closer to that inferred from the observed gamma-ray luminosity.
At early times, the shocked plasma must have had much higher density,
which should have resulted in an X-ray luminosity close to the shock
power, yet no novae to date have been observed to have such high
luminosity X-ray emissions.
\end{itemize}

\section*{Acknowledgments}

We thank the support of the late Neil Gehrels and
the \swift\ mission team for their help in scheduling the observations
presented in this article.  This research has made use of data and
software provided by the High Energy Astrophysics Science Archive Research
Center (HEASARC), which is a service of the Astrophysics Science Division
at NASA/GSFC and the High Energy Astrophysics Division of the Smithsonian
Astrophysical Observatory. The scientific results reported in this article
are based in part on observations made by the \chandra\ X-ray Observatory.
They are also based in part on data obtained from the \suzaku\ satellite,
a collaborative mission between the space agencies of Japan (JAXA) and
the USA (NASA). L.C. is grateful for support from a Cottrell scholarship
of the Research Corporation for Science Advancement, NSF grant AST-1751874,
and NASA GI programs Fermi-NNX14AQ36G, Fermi-80NSSC18K1746, and
NuSTAR-80NSSC19K0522. J.L.S. is grateful for support from
Hubble Space Telescope General Observer grants HST-GO-13796 and
HST-GO-15438 from NASA, and from NSF grant NSF AST-1816100.

\section*{Data Availability}

The data presented in this article are all publicly available at the
High Energy Astrophysics Science Archive Research Center
(https://heasarc.gsfc.nasa.gov).




\bibliography{v959mon}

\newcommand{\noop}[1]{}
\begin{thebibliography}{}
\makeatletter
\relax
\def\mn@urlcharsother{\let\do\@makeother \do\$\do\&\do\#\do\^\do\_\do\%\do\~}
\def\mn@doi{\begingroup\mn@urlcharsother \@ifnextchar [ {\mn@doi@}
  {\mn@doi@[]}}
\def\mn@doi@[#1]#2{\def\@tempa{#1}\ifx\@tempa\@empty \href
  {http://dx.doi.org/#2} {doi:#2}\else \href {http://dx.doi.org/#2} {#1}\fi
  \endgroup}
\def\mn@eprint#1#2{\mn@eprint@#1:#2::\@nil}
\def\mn@eprint@arXiv#1{\href {http://arxiv.org/abs/#1} {{\tt arXiv:#1}}}
\def\mn@eprint@dblp#1{\href {http://dblp.uni-trier.de/rec/bibtex/#1.xml}
  {dblp:#1}}
\def\mn@eprint@#1:#2:#3:#4\@nil{\def\@tempa {#1}\def\@tempb {#2}\def\@tempc
  {#3}\ifx \@tempc \@empty \let \@tempc \@tempb \let \@tempb \@tempa \fi \ifx
  \@tempb \@empty \def\@tempb {arXiv}\fi \@ifundefined
  {mn@eprint@\@tempb}{\@tempb:\@tempc}{\expandafter \expandafter \csname
  mn@eprint@\@tempb\endcsname \expandafter{\@tempc}}}

\bibitem[\protect\citeauthoryear{{Ackermann}, {Ajello}, {Albert}
  et~al.}{{Ackermann} et~al.}{2014}]{Ackermann14}
{Ackermann} M.,  {Ajello} M.,  {Albert} A.,   et~al., 2014, \mn@doi [Science]
  {10.1126/science.1253947}, \href
  {http://adsabs.harvard.edu/abs/2014Sci...345..554A} {345, 554}

\bibitem[\protect\citeauthoryear{{Aydi} et~al.,}{{Aydi} et~al.}{2020}]{Aydi20}
{Aydi} E.,  et~al., 2020, \mn@doi [Nature Astronomy]
  {10.1038/s41550-020-1070-y}, \href
  {https://ui.adsabs.harvard.edu/abs/2020NatAs...4..776A} {4, 776}

\bibitem[\protect\citeauthoryear{{Balman}, {Krautter}  \&
  {{\"O}gelman}}{{Balman} et~al.}{1998}]{Balman98}
{Balman} {\c S}.,  {Krautter} J.,   {{\"O}gelman} H.,  1998, \mn@doi [\apj]
  {10.1086/305600}, \href {http://adsabs.harvard.edu/abs/1998ApJ...499..395B}
  {499, 395}

\bibitem[\protect\citeauthoryear{{Bode} \& {Evans}}{{Bode} \&
  {Evans}}{2008}]{CNII}
{Bode} M.,  {Evans} A.,  eds, 2008, {Classical Novae, 2nd Edition}.
Cambridge University Press, Cambridge, UK

\bibitem[\protect\citeauthoryear{{Bode}, {O'Brien}, {Osborne}  et~al.}{{Bode}
  et~al.}{2006}]{Bode06}
{Bode} M.~F.,  {O'Brien} T.~J.,  {Osborne} J.~P.,   et~al., 2006, \mn@doi
  [\apj] {10.1086/507980}, \href
  {http://adsabs.harvard.edu/abs/2006ApJ...652..629B} {652, 629}

\bibitem[\protect\citeauthoryear{{Cash}}{{Cash}}{1979}]{Cash79}
{Cash} W.,  1979, \mn@doi [\apj] {10.1086/156922}, \href
  {http://adsabs.harvard.edu/abs/1979ApJ...228..939C} {228, 939}

\bibitem[\protect\citeauthoryear{{Cheung}, {Hays}, {Venters}, {Donato}  \&
  {Corbet}}{{Cheung} et~al.}{2012a}]{Cheung12a}
{Cheung} C.~C.,  {Hays} E.,  {Venters} T.,  {Donato} D.,   {Corbet} R.~H.~D.,
  2012a, The Astronomer's Telegram, \href
  {http://adsabs.harvard.edu/abs/2012ATel.4224....1C} {4224}

\bibitem[\protect\citeauthoryear{{Cheung}, {Shore}, {De Gennaro Aquino},
  {Charbonnel}, {Edlin}, {Hays}, {Corbet}  \& {Wood}}{{Cheung}
  et~al.}{2012b}]{Cheung12b}
{Cheung} C.~C.,  {Shore} S.~N.,  {De Gennaro Aquino} I.,  {Charbonnel} S.,
  {Edlin} J.,  {Hays} E.,  {Corbet} R.~H.~D.,   {Wood} D.~L.,  2012b, The
  Astronomer's Telegram, \href
  {http://adsabs.harvard.edu/abs/2012ATel.4310....1C} {4310}

\bibitem[\protect\citeauthoryear{{Cheung}, {Jean}, {Shore}  et~al.}{{Cheung}
  et~al.}{2016}]{Cheung16}
{Cheung} C.~C.,  {Jean} P.,  {Shore} S.~N.,   et~al., 2016, \mn@doi [\apj]
  {10.3847/0004-637X/826/2/142}, \href
  {http://adsabs.harvard.edu/abs/2016ApJ...826..142C} {826, 142}

\bibitem[\protect\citeauthoryear{{Chomiuk}, {Linford}, {Yang}
  et~al.}{{Chomiuk} et~al.}{2014a}]{Chomiuk14a}
{Chomiuk} L.,  {Linford} J.~D.,  {Yang} J.,   et~al., 2014a, \mn@doi [\nat]
  {10.1038/nature13773}, \href
  {http://adsabs.harvard.edu/abs/2014Natur.514..339C} {514, 339}

\bibitem[\protect\citeauthoryear{{Chomiuk}, {Nelson}, {Mukai}
  et~al.}{{Chomiuk} et~al.}{2014b}]{Chomiuk14b}
{Chomiuk} L.,  {Nelson} T.,  {Mukai} K.,   et~al., 2014b, \mn@doi [\apj]
  {10.1088/0004-637X/788/2/130}, \href
  {http://adsabs.harvard.edu/abs/2014ApJ...788..130C} {788, 130}

\bibitem[\protect\citeauthoryear{{Diaz}, {Williams}, {Luna}, {Moraes}  \&
  {Takeda}}{{Diaz} et~al.}{2010}]{Diaz10}
{Diaz} M.~P.,  {Williams} R.~E.,  {Luna} G.~J.,  {Moraes} M.,   {Takeda} L.,
  2010, \mn@doi [\aj] {10.1088/0004-6256/140/6/1860}, \href
  {http://adsabs.harvard.edu/abs/2010AJ....140.1860D} {140, 1860}

\bibitem[\protect\citeauthoryear{{Evans} et~al.,}{{Evans}
  et~al.}{2014}]{Evans14}
{Evans} P.~A.,  et~al., 2014, \mn@doi [\apjs] {10.1088/0067-0049/210/1/8},
  \href {http://adsabs.harvard.edu/abs/2014ApJS..210....8E} {210, 8}

\bibitem[\protect\citeauthoryear{{Fujikawa}, {Yamaoka}  \& {Nakano}}{{Fujikawa}
  et~al.}{2012}]{Fujikawa12}
{Fujikawa} S.,  {Yamaoka} H.,   {Nakano} S.,  2012, Central Bureau Electronic
  Telegrams, \href {http://adsabs.harvard.edu/abs/2012CBET.3202....1F} {3202}

\bibitem[\protect\citeauthoryear{{G{\"u}ver} \& {{\"O}zel}}{{G{\"u}ver} \&
  {{\"O}zel}}{2009}]{Guver09}
{G{\"u}ver} T.,  {{\"O}zel} F.,  2009, \mn@doi [\mnras]
  {10.1111/j.1365-2966.2009.15598.x}, \href
  {http://adsabs.harvard.edu/abs/2009MNRAS.400.2050G} {400, 2050}

\bibitem[\protect\citeauthoryear{{Healy}, {O'Brien}, {Beswick}, {Avison}  \&
  {Argo}}{{Healy} et~al.}{2017}]{Healy17}
{Healy} F.,  {O'Brien} T.~J.,  {Beswick} R.,  {Avison} A.,   {Argo} M.~K.,
  2017, \mn@doi [\mnras] {10.1093/mnras/stx1087}, \href
  {http://adsabs.harvard.edu/abs/2017MNRAS.469.3976H} {469, 3976}

\bibitem[\protect\citeauthoryear{{Helton} et~al.,}{{Helton}
  et~al.}{2012}]{Helton12}
{Helton} L.~A.,  et~al., 2012, \mn@doi [\apj] {10.1088/0004-637X/755/1/37},
  \href {http://adsabs.harvard.edu/abs/2012ApJ...755...37H} {755, 37}

\bibitem[\protect\citeauthoryear{{Ishida}, {Morio}  \& {Ueda}}{{Ishida}
  et~al.}{2004}]{Ishida04}
{Ishida} M.,  {Morio} K.,   {Ueda} Y.,  2004, \mn@doi [\apj] {10.1086/380780},
  \href {http://adsabs.harvard.edu/abs/2004ApJ...601.1088I} {601, 1088}

\bibitem[\protect\citeauthoryear{{Li}, {Kong}, {Charles}  et~al.}{{Li}
  et~al.}{2012}]{Li12}
{Li} K.~L.,  {Kong} A.~K.~H.,  {Charles} P.~A.,   et~al., 2012, \mn@doi [\apj]
  {10.1088/0004-637X/761/2/99}, \href
  {http://adsabs.harvard.edu/abs/2012ApJ...761...99L} {761, 99}

\bibitem[\protect\citeauthoryear{{Linford}, {Ribeiro}, {Chomiuk}
  et~al.}{{Linford} et~al.}{2015}]{Linford15}
{Linford} J.~D.,  {Ribeiro} V.~A.~R.~M.,  {Chomiuk} L.,   et~al., 2015, \mn@doi
  [\apj] {10.1088/0004-637X/805/2/136}, \href
  {http://adsabs.harvard.edu/abs/2015ApJ...805..136L} {805, 136}

\bibitem[\protect\citeauthoryear{{Metzger}, {Hasco{\"e}t}, {Vurm},
  {Beloborodov}, {Chomiuk}, {Sokoloski}  \& {Nelson}}{{Metzger}
  et~al.}{2014}]{Metzger14}
{Metzger} B.~D.,  {Hasco{\"e}t} R.,  {Vurm} I.,  {Beloborodov} A.~M.,
  {Chomiuk} L.,  {Sokoloski} J.~L.,   {Nelson} T.,  2014, \mn@doi [\mnras]
  {10.1093/mnras/stu844}, \href
  {http://adsabs.harvard.edu/abs/2014MNRAS.442..713M} {442, 713}

\bibitem[\protect\citeauthoryear{{Mukai} \& {Ishida}}{{Mukai} \&
  {Ishida}}{2001}]{Mukai01}
{Mukai} K.,  {Ishida} M.,  2001, \mn@doi [\apj] {10.1086/320220}, \href
  {http://adsabs.harvard.edu/abs/2001ApJ...551.1024M} {551, 1024}

\bibitem[\protect\citeauthoryear{{Mukai}, {Orio}  \& {Della Valle}}{{Mukai}
  et~al.}{2008}]{Mukai08}
{Mukai} K.,  {Orio} M.,   {Della Valle} M.,  2008, \mn@doi [\apj]
  {10.1086/529362}, \href {http://adsabs.harvard.edu/abs/2008ApJ...677.1248M}
  {677, 1248}

\bibitem[\protect\citeauthoryear{{Munari}, {Dallaporta}, {Castellani}
  et~al.}{{Munari} et~al.}{2013}]{Munari13}
{Munari} U.,  {Dallaporta} S.,  {Castellani} F.,   et~al., 2013, \mn@doi
  [\mnras] {10.1093/mnras/stt1340}, \href
  {http://adsabs.harvard.edu/abs/2013MNRAS.435..771M} {435, 771}

\bibitem[\protect\citeauthoryear{{Nelson}, {Donato}, {Mukai}, {Sokoloski}  \&
  {Chomiuk}}{{Nelson} et~al.}{2012}]{Nelson12}
{Nelson} T.,  {Donato} D.,  {Mukai} K.,  {Sokoloski} J.,   {Chomiuk} L.,  2012,
  \mn@doi [\apj] {10.1088/0004-637X/748/1/43}, \href
  {http://adsabs.harvard.edu/abs/2012ApJ...748...43N} {748, 43}

\bibitem[\protect\citeauthoryear{{Nelson}, {Chomiuk}, {Roy}  et~al.}{{Nelson}
  et~al.}{2014}]{Nelson14}
{Nelson} T.,  {Chomiuk} L.,  {Roy} N.,   et~al., 2014, \mn@doi [\apj]
  {10.1088/0004-637X/785/1/78}, \href
  {http://adsabs.harvard.edu/abs/2014ApJ...785...78N} {785, 78}

\bibitem[\protect\citeauthoryear{{Nelson}, {Mukai}, {Li}  et~al.}{{Nelson}
  et~al.}{2019}]{Nelson19}
{Nelson} T.,  {Mukai} K.,  {Li} K.-L.,   et~al., 2019, \mn@doi [\apj]
  {10.3847/1538-4357/aafb6d}, \href
  {http://adsabs.harvard.edu/abs/2019ApJ...872...86N} {872, 86}

\bibitem[\protect\citeauthoryear{{Ness}, {Schwarz}, {Retter}  et~al.}{{Ness}
  et~al.}{2007}]{Ness07}
{Ness} J.-U.,  {Schwarz} G.~J.,  {Retter} A.,   et~al., 2007, \mn@doi [\apj]
  {10.1086/518084}, \href {http://adsabs.harvard.edu/abs/2007ApJ...663..505N}
  {663, 505}

\bibitem[\protect\citeauthoryear{{Ness}, {Shore}, {Drake}, {Osborne}, {Page},
  {Beardmore}, {Schwarz}  \& {Starrfield}}{{Ness} et~al.}{2012}]{Ness12}
{Ness} J.-U.,  {Shore} S.~N.,  {Drake} J.~J.,  {Osborne} J.~P.,  {Page} K.~L.,
  {Beardmore} A.,  {Schwarz} G.,   {Starrfield} S.,  2012, The Astronomer's
  Telegram, \href {http://adsabs.harvard.edu/abs/2012ATel.4569....1N} {4569, 1}

\bibitem[\protect\citeauthoryear{Newville, Stensitzki, Allen  \&
  Ingargiola}{Newville et~al.}{2014}]{Newville14}
Newville M.,  Stensitzki T.,  Allen D.~B.,   Ingargiola A.,  2014, {LMFIT:
  Non-Linear Least-Square Minimization and Curve-Fitting for Python},
  \mn@doi{10.5281/zenodo.11813}, \url {https://doi.org/10.5281/zenodo.11813}

\bibitem[\protect\citeauthoryear{{O'Brien}, {Lloyd}  \& {Bode}}{{O'Brien}
  et~al.}{1994}]{O'Brien94}
{O'Brien} T.~J.,  {Lloyd} H.~M.,   {Bode} M.~F.,  1994, \mnras, \href
  {http://adsabs.harvard.edu/abs/1994MNRAS.271..155O} {271, 155}

\bibitem[\protect\citeauthoryear{{Orio}, {Rana}, {Page}, {Sokoloski}  \&
  {Harrison}}{{Orio} et~al.}{2015}]{Orio15}
{Orio} M.,  {Rana} V.,  {Page} K.~L.,  {Sokoloski} J.,   {Harrison} F.,  2015,
  \mn@doi [\mnras] {10.1093/mnrasl/slu195}, \href
  {http://adsabs.harvard.edu/abs/2015MNRAS.448L..35O} {448, L35}

\bibitem[\protect\citeauthoryear{{Page}, {Osborne}, {Wagner}  et~al.}{{Page}
  et~al.}{2013}]{Page13}
{Page} K.~L.,  {Osborne} J.~P.,  {Wagner} R.~M.,   et~al., 2013, \mn@doi
  [\apjl] {10.1088/2041-8205/768/2/L26}, \href
  {http://adsabs.harvard.edu/abs/2013ApJ...768L..26P} {768, L26}

\bibitem[\protect\citeauthoryear{{Pejcha}, {Metzger}  \& {Tomida}}{{Pejcha}
  et~al.}{2016}]{Pejcha16}
{Pejcha} O.,  {Metzger} B.~D.,   {Tomida} K.,  2016, \mn@doi [\mnras]
  {10.1093/mnras/stw1481}, \href
  {https://ui.adsabs.harvard.edu/abs/2016MNRAS.461.2527P} {461, 2527}

\bibitem[\protect\citeauthoryear{{Peretz}, {Orio}, {Behar}, {Bianchini},
  {Gallagher}, {Rauch}, {Tofflemire}  \& {Zemko}}{{Peretz}
  et~al.}{2016}]{Peretz16}
{Peretz} U.,  {Orio} M.,  {Behar} E.,  {Bianchini} A.,  {Gallagher} J.,
  {Rauch} T.,  {Tofflemire} B.,   {Zemko} P.,  2016, \apj, \href
  {http://adsabs.harvard.edu/abs/2016ApJ...829....2P/abstract} {829, 2}

\bibitem[\protect\citeauthoryear{{Porquet}, {Mewe}, {Dubau}, {Raassen}  \&
  {Kaastra}}{{Porquet} et~al.}{2001}]{Porquet01}
{Porquet} D.,  {Mewe} R.,  {Dubau} J.,  {Raassen} A.~J.~J.,   {Kaastra} J.~S.,
  2001, \mn@doi [\aap] {10.1051/0004-6361:20010959}, \href
  {http://adsabs.harvard.edu/abs/2001A%26A...376.1113P} {376, 1113}

\bibitem[\protect\citeauthoryear{{Ribeiro}, {Munari}  \& {Valisa}}{{Ribeiro}
  et~al.}{2013}]{Ribeiro13}
{Ribeiro} V.~A.~R.~M.,  {Munari} U.,   {Valisa} P.,  2013, \mn@doi [\apj]
  {10.1088/0004-637X/768/1/49}, \href
  {http://adsabs.harvard.edu/abs/2013ApJ...768...49R} {768, 49}

\bibitem[\protect\citeauthoryear{{Schwarz}, {Ness}, {Osborne}
  et~al.}{{Schwarz} et~al.}{2011}]{Schwarz11}
{Schwarz} G.~J.,  {Ness} J.-U.,  {Osborne} J.~P.,   et~al., 2011, \mn@doi
  [\apjs] {10.1088/0067-0049/197/2/31}, \href
  {http://adsabs.harvard.edu/abs/2011ApJS..197...31S} {197, 31}

\bibitem[\protect\citeauthoryear{{Seaquist} \& {Palimaka}}{{Seaquist} \&
  {Palimaka}}{1977}]{Seaquist77}
{Seaquist} E.~R.,  {Palimaka} J.,  1977, \mn@doi [\apj] {10.1086/155625}, \href
  {http://adsabs.harvard.edu/abs/1977ApJ...217..781S} {217, 781}

\bibitem[\protect\citeauthoryear{{Shore}, {De Gennaro Aquino}, {Schwarz},
  {Augusteijn}, {Cheung}, {Walter}  \& {Starrfield}}{{Shore}
  et~al.}{2013}]{Shore13}
{Shore} S.~N.,  {De Gennaro Aquino} I.,  {Schwarz} G.~J.,  {Augusteijn} T.,
  {Cheung} C.~C.,  {Walter} F.~M.,   {Starrfield} S.,  2013, \mn@doi [\aap]
  {10.1051/0004-6361/201321095}, \href
  {http://adsabs.harvard.edu/abs/2013A%26A...553A.123S} {553, A123}

\bibitem[\protect\citeauthoryear{{Smith} \& {Hughes}}{{Smith} \&
  {Hughes}}{2010}]{Smith10}
{Smith} R.~K.,  {Hughes} J.~P.,  2010, \mn@doi [\apj]
  {10.1088/0004-637X/718/1/583}, \href
  {http://adsabs.harvard.edu/abs/2010ApJ...718..583S} {718, 583}

\bibitem[\protect\citeauthoryear{{Sokoloski}, {Luna}, {Mukai}  \&
  {Kenyon}}{{Sokoloski} et~al.}{2006}]{Sokoloski06}
{Sokoloski} J.~L.,  {Luna} G.~J.~M.,  {Mukai} K.,   {Kenyon} S.~J.,  2006,
  \mn@doi [\nat] {10.1038/nature04893}, \href
  {http://adsabs.harvard.edu/abs/2006Natur.442..276S} {442, 276}

\bibitem[\protect\citeauthoryear{{Sokoloski}, {Lawrence}, {Crotts}  \&
  {Mukai}}{{Sokoloski} et~al.}{2017}]{Sokoloski17}
{Sokoloski} J.~L.,  {Lawrence} S.,  {Crotts} A. P.~S.,   {Mukai} K.,  2017,
  arXiv:1702.05898, \href
  {https://ui.adsabs.harvard.edu/abs/2017arXiv170205898S} {}

\bibitem[\protect\citeauthoryear{{Sokolovsky} et~al.,}{{Sokolovsky}
  et~al.}{2020}]{Sokolovsky20}
{Sokolovsky} K.~V.,  et~al., 2020, \mn@doi [\mnras] {10.1093/mnras/staa2104},
  \href {https://ui.adsabs.harvard.edu/abs/2020MNRAS.497.2569S} {497, 2569}

\bibitem[\protect\citeauthoryear{{Tarasova}}{{Tarasova}}{2014}]{Tarasova14}
{Tarasova} T.~N.,  2014, \mn@doi [Astronomy Letters]
  {10.1134/S1063773714050053}, \href
  {http://adsabs.harvard.edu/abs/2014AstL...40..309T} {40, 309}

\bibitem[\protect\citeauthoryear{{Verner}, {Ferland}, {Korista}  \&
  {Yakovlev}}{{Verner} et~al.}{1996}]{Verner96}
{Verner} D.~A.,  {Ferland} G.~J.,  {Korista} K.~T.,   {Yakovlev} D.~G.,  1996,
  \mn@doi [\apj] {10.1086/177435}, \href
  {http://adsabs.harvard.edu/abs/1996ApJ...465..487V} {465, 487}

\bibitem[\protect\citeauthoryear{{Weston} et~al.,}{{Weston}
  et~al.}{2016}]{Weston16}
{Weston} J.~H.~S.,  et~al., 2016, \mn@doi [\mnras] {10.1093/mnras/stv3019},
  \href {http://adsabs.harvard.edu/abs/2016MNRAS.457..887W} {457, 887}

\bibitem[\protect\citeauthoryear{{Wilms}, {Allen}  \& {McCray}}{{Wilms}
  et~al.}{2000}]{Wilms00}
{Wilms} J.,  {Allen} A.,   {McCray} R.,  2000, \mn@doi [\apj] {10.1086/317016},
  \href {http://adsabs.harvard.edu/abs/2000ApJ...542..914W} {542, 914}

\makeatother
\end{thebibliography}



\bsp	
\label{lastpage}
\end{document}